\documentclass[12pt,oneside,a4paper,english]{article}
\usepackage[T1]{fontenc}
\usepackage[utf8]{inputenc}
\usepackage[margin=2.25cm,headheight=26pt,includeheadfoot]{geometry}
\usepackage{caption}
\usepackage[english]{babel}
\usepackage{listings}
\usepackage{color}
\usepackage{titlesec}
\usepackage{titling}
\usepackage{changepage}
\usepackage{amsmath}
\usepackage{hyperref}
\usepackage{enumitem}
\usepackage{graphicx}
\usepackage[font=normalsize, labelfont=bf]{caption}
\usepackage{subcaption}
\usepackage[export]{adjustbox}
\usepackage{caption}
\usepackage{fancyhdr}
\usepackage{enumitem}
\usepackage{lastpage}
\usepackage{caption}
\usepackage{tocloft}
\usepackage{setspace}
\usepackage{multirow}
\usepackage{titling}
\usepackage{float}
\usepackage{comment}
\usepackage{booktabs}
\usepackage{indentfirst}
\usepackage{lscape}
\usepackage{booktabs,caption}
\usepackage[flushleft]{threeparttable}
\usepackage[english]{nomencl}
\usepackage{xcolor}

\title{BIOREME Study Group Report} 
\makeatletter\let\Title\@title\makeatother

\newlist{steps}{enumerate}{1}
\setlist[steps, 1]{leftmargin=1.5cm,label = Step \arabic*:}

\begin{document}
\pagenumbering{roman} 

\pdfoutput=1
\begin{titlepage}
\begin{center}
\vspace{1.0cm}
\includegraphics[width=0.4\textwidth]{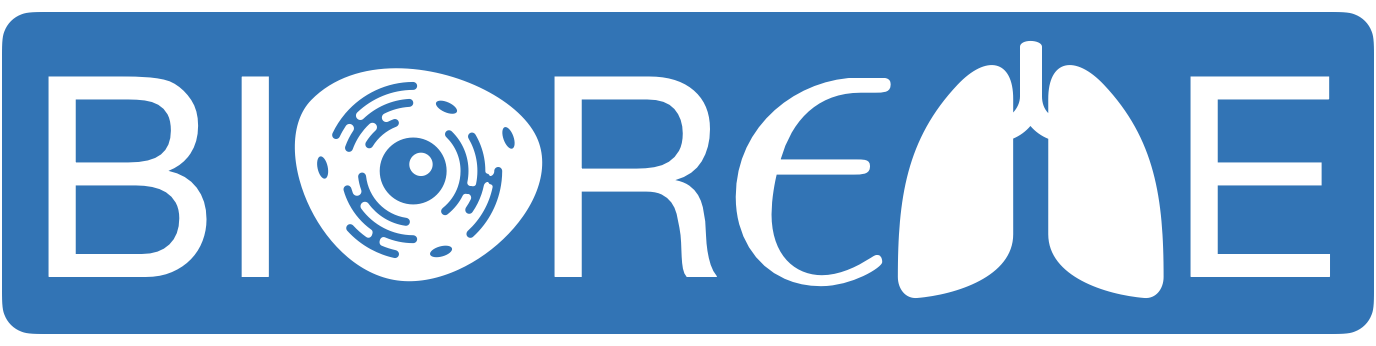}~\\[1cm]

{\large\textbf{A Mathematical Study Group in Data-Driven Biophysical Modelling for Respiratory Diseases with Industry and Clinicians}}
\vspace{.5cm}

\vspace{2.0cm}

\hrule
\vspace{.5cm}
{\huge\textbf{Can audio recordings be used to detect leaks and coughs during mechanical insufflation exsufflation (MI-E) treatment?}} 
\vspace{.5cm}

\hrule
\vspace{1.5cm}

\textsc{\textbf{Authors}}\\
\vspace{.5cm}
\centering


Jonathan Brady -- University of Manchester\\
Vijay Chandiramani -- University of Bristol\\
Michelle Chatwin -- Royal Brompton and Harefield NHS Foundation Trust\\
Emmanuel Lwele -- Sheffield Hallam University\\
Aminat Yetunde Saula -- University of Bath\\
Peter Stewart -- University of Glasgow\\
Toby Stokes-- BREAS Medical\\

\vspace{1cm}
*Corresponding author: 
General enquiries: \href{mailto:contact@bioreme.net}{contact@bioreme.net}

\vspace{2cm}

\centering Date of publication: \today 
\end{center}
\end{titlepage}

\newpage
\doublespacing
\renewcommand{\baselinestretch}{1}\normalsize
\tableofcontents
\renewcommand{\baselinestretch}{1}\normalsize
\thispagestyle{fancy} 

\newpage
\pagenumbering{arabic} 
\fancyfoot[C]{Page \thepage\ of \pageref{EndOfText}}

\section{Introduction} 
\label{sec:intro}

This report relates to a study group hosted by the EPSRC funded network, Integrating data-driven BIOphysical models into REspiratory MEdicine \href{https://www.bioreme.net/about}{(BIOREME)}, and supported by \href{https://www.softmech.org/}{SofTMech}  and \href{https://iuk.ktn-uk.org/industrial-maths/}{Innovate UK, Business Connect}. The BIOREME network hosts events, including this study group, to bring together multi-disciplinary researchers, clinicians, companies and charities to catalyse research in the applications of mathematical modelling for respiratory medicine. The goal of this study group was to provide an interface between companies, clinicians, and mathematicians to develop mathematical tools to the problems presented. The study group was held at The University of Glasgow on the 17 - 21 June 2024 and was attended by 16 participants from 8 different institutions. Below details the technical report of one of the challenges and the methods developed by the team of researchers who worked on this challenge.

\subsection{The Challenge}
\label{sec:challenge}

\textbf{Background:} 
The mechanical insufflation-exsufflation (MI-E) device, manufactured by a global company  helps  people with neuromuscular weakness (pwNMW) to cough more effectively. An ineffective cough can lead to a build up of secretions in the respiratory tract. This is a major cause of mortality and morbidity in patients with weak cough, typically those with neuromuscular disease due to weak inspiratory of expiratory muscles or bulbar 
impairment.  Three steps make up a cough: inhalation, increased pressure in the throat and lungs caused by the vocal cords closing, and finally, an explosive release of air caused by the vocal cords opening, which produces the cough`s distinctive sound. The MI-E device emulates the cough procedure by providing a deep breath in (insufflation) followed by a rapid switch to negative pressure (exsufflation). The simulated changes in flow from the device mimic what occurs naturally during a cough and thereby assist cough strength, avoiding build-up of secretions in the lungs.

\textbf{The problem:} A clinician/carer can audibly detect: 
\begin{itemize}
    \item Switch between positive and negative pressure
    \item Any resulting cough
    \item Device leakiness
\end{itemize}

For the MI-E device to be effective, there needs to be minimal leak. Leak is a problem because the inspiratory volume cannot be achieved, or it can only be achieved over a prolonged period. This can lead to ineffective treatment and discomfort for the person using the device. One way to evaluate leak is via the pressure profile. However, not all devices can provide both the pressure and flow signals, which can complicate evaluation for some clinicians. The audibility of a patient`s cough can be used to determine its effectiveness. The patient is asked to cough and depending on the strength and sound of the cough, it is deemed effective or ineffective. Leak is often audible and therefore an analysis of the sound generated may be one option to evaluate efficacy.  

\textbf{Research question}
This study, therefore, is focused on addressing the following research question:

\emph{Can a machine identify suboptimal MI-E treatments through analysis of an audio recording by detecting the presence of leak or the lack of a patient cough?}

This will help us to understand options for supporting users of our devices outside of medical settings.

\textbf{Data available:} The manufacturer has made available a dataset of 81 treatment audio recordings 
of MI-E which have been labelled to indicate the presence of cough and leak. The recordings are of real patients being treated with a range of MI-E devices. The device was demonstrated to the study group along with videos of patients being treated and provided a better understanding of how it functions and the problem presented.

\subsection{Proposed Solution}
\label{sec:solution}
The study group proposed different solutions to the address the research question as follows: 
\begin{enumerate}
\item Signal processing
\item Denoising
\item Machine Learning
\end{enumerate}
Each Solution is summarised in the sections that follow.
\subsection{Solution 1 - Adaptive Neuro-Fuzzy Inference System (ANFIS)}
\subsubsection*{\textbf{Introduction}}
The classification of respiratory sounds is critical for monitoring and assessing respiratory conditions, especially when using devices like Mechanical Insufflation-Exsufflation (MI-E). The proposed solution combines Mel Frequency Cepstral Coefficients (MFCCs) and Melspectrograms to create a new feature set for classifying respiratory sounds. By integrating these two features, the proposed method captures both short-term spectral details and long-term temporal patterns, thus improving the detection of abnormalities such as coughs and leaks.

The combined feature set is used to train an Adaptive Neuro-Fuzzy Inference System (ANFIS), which integrates fuzzy logic and neural network principles to classify respiratory sounds effectively. The model’s performance is evaluated using a 5-fold cross-validation approach, ensuring robust results and generalization across different subsets of the dataset.

\subsection{Solution 2 - Wavelet Denoising}
The proposed approach employs wavelet denoising to enhance the audio signals from the provided file. This technique effectively reduces noise while preserving the key features of the signal. Specifically, the audio signal \( x[n] \) is decomposed using the `sym4' wavelet at level 5, producing a set of wavelet coefficients \( C = \{c_{j,k}\} \).

Traditional wavelet denoising methods focus on retaining approximations while thresholding the detail coefficients. The threshold value is typically determined based on the noise intensity, which can vary across different decomposition levels. Since real audio signals often contain non-stationary noise, accurately estimating noise intensity requires accounting for variations in standard deviation across these levels \cite{baldazzi2020systematic}. To address this, the Median Absolute Deviation (MAD) is used to estimate the standard deviation of a normally distributed signal by scaling MAD with a constant. In wavelet denoising, MAD provides a robust estimate of noise intensity by focusing on the smallest wavelet coefficients. Specifically, the noise level \(\sigma\) is estimated as follows \cite{baldazzi2020systematic, verma2012performance}:

\[
\sigma = \frac{\text{median}\left(\left\{ |c_{j,k}| \right\}\right)}{0.6745},
\]

where \(c_{j,k}\) are the wavelet coefficients at a given decomposition level \(j\) and time index \(k\). The constant \(0.6745\) scales MAD to approximate the standard deviation of a normal distribution, ensuring accurate noise level estimation.

To denoise the signal, a soft thresholding function \( T_{\lambda}(c_{j,k}) \) is applied to the wavelet coefficients. The threshold \( \lambda \) is determined by the following equation \cite{donoho1994ideal,donoho1995noising}:

\[
\lambda = \alpha \cdot \sigma \cdot \sqrt{2 \log n},
\]

where \( \alpha \) is a scaling factor, \( \sigma \) is the estimated noise level, and \( n \) is the length of the audio signal. The denoised audio signal \( \tilde{x}[n] \) is then reconstructed using the inverse wavelet transform:

\[
\tilde{x}[n] = \sum_{j,k} T_{\lambda}(c_{j,k}) \cdot \psi_{j,k}[n],
\]

where \( \psi_{j,k}[n] \) denotes the wavelet basis functions in discrete time. The resulting denoised audio signal \( \tilde{x}[n] \) is subsequently utilised for spectral analysis in \autoref{sec:method 2}.
\subsection{Solution 3 - Machine Learning}
After reviewing the audio signals and analysing them to detect patterns, Machine Learning provides a promising approach to address the research question of detecting coughs from audio signals. To start with, it is pertinent to explain the concept of Machine Learning in the context of cough detection. Machine learning (ML) is a subset of artificial intelligence (AI) that involves the development of algorithms and statistical models that enable computers to perform tasks without explicit instructions \cite{murphy2012machine}. Instead of being programmed with specific rules, ML systems learn patterns and relationships from data, improving their performance on tasks over time as they are exposed to more data. Since the occurrence of cough events is usually accompanied by the production of some specific sounds, researchers have been able to detect cough signals by analysing the characteristics of the audio signal. Alsabek et al.\cite{alsabek2020studying} highlighted the significance of extracting Mel Frequency Cepstral Coefficients (MFCC) from COVID-19 and non-COVID-19 samples during cough signal processing. Monge-Alvarez et al. \cite{monge2018robust} demonstrated the effectiveness of using local Hu moments as robust features for cough detection through audio signals.  Pramono et al. \cite{pramono2019automatic} created an algorithm based on audio signal features and logistic regression to identify cough events. After study of the above research and considering the context of the research question, the Machine Learning solution was devised as per the Method defined in Section \ref{sec:method 3}.

\section{Methods} 
\label{sec:methods}

\subsection{Method 1- Adaptive Neuro-Fuzzy Inference System (ANFIS)}\label{sec:}
The Adaptive Neuro-Fuzzy Inference System (ANFIS) is a hybrid artificial intelligence model that combines the strengths of both neural networks and fuzzy logic principles \cite{walia2015anfis}. It was introduced to take advantage of the learning capabilities of neural networks and the interpretability of fuzzy logic, enabling the system to model complex, nonlinear relationships. ANFIS is particularly useful in situations where uncertainty or vague information is present, making it an ideal choice for tasks such as classification, function approximation, and pattern recognition \cite{walia2015anfis}.
The Adaptive Neuro-Fuzzy Inference System (ANFIS) was utilized to detect coughs in the denoised audio signals using the combined feature set. This methodology consists of the following key stages: gathering data, preparing the data for analysis, selecting a suitable model, and evaluating the model’s performance.

\subsubsection {Gathering Data}
The dataset comprises 81 audio recordings from MI-E treatments, with labels indicating the presence or absence of cough and leak sounds. 

\begin{figure}[H]
    \centering
    \includegraphics[width=0.9\textwidth]{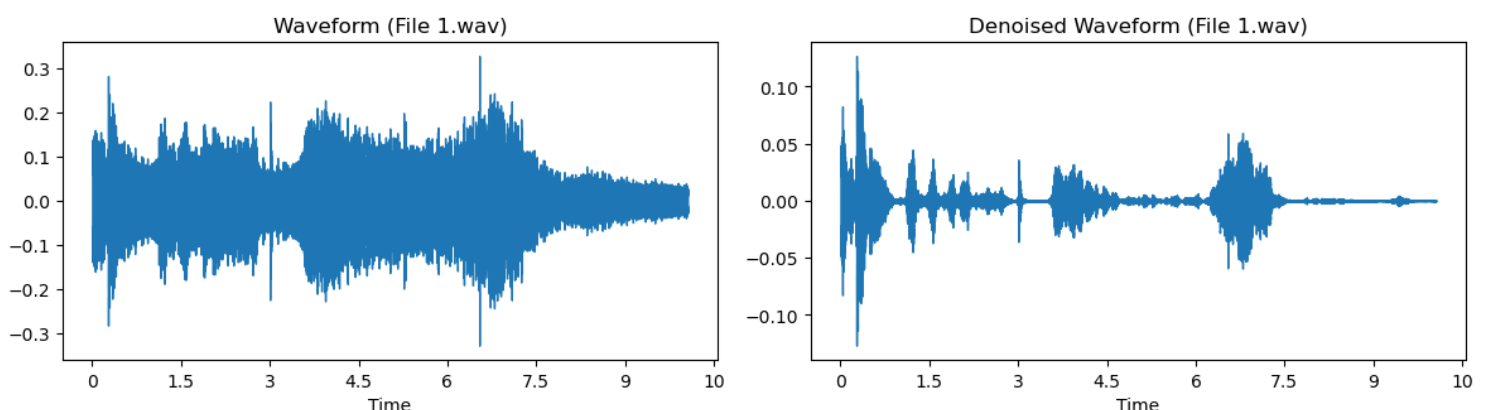}
    \caption{Comparison of the original and denoised audio waveforms for File 1.wav. The left plot shows the original waveform, which has a higher noise level, while the right plot displays the denoised waveform, showing a clearer signal with reduced background noise.}
    \label{fig:waveform_comparison}
\end{figure}

The dataset is diverse, encompassing a range of real-world conditions, including varied noise levels and patient respiratory activities. The labels serve as the ground truth for training and evaluating the model, making it suitable for testing the robustness of machine learning models in respiratory sound classification. Quality checks were performed on each recording to ensure that key sound events were distinguishable. The signals shown in Figure \ref{fig:waveform_comparison}, compares the original and denoised waveforms for the recorded audio.

\subsubsection {Preparing the Data}
\noindent \textbf{1. Noise Reduction:}
Spectral gating was used to reduce background noise (denoise) by estimating the noise profile and subtracting it from the original signal. This step ensured that the features extracted would represent core respiratory sound characteristics, such as coughs and leaks.

\noindent \textbf{2. Signal Segmentation using STFT:}
The Short-Time Fourier Transform (STFT) was applied to the audio signals to segment them into frames, each capturing a small segment of the signal and allowing for a time-frequency analysis. This segmentation is defined as:

\[
\text{STFT}\{{x}[n]\}(m, \omega) = \sum_{n=0}^{N-1} \tilde{x}[m + n] \cdot w[n] \cdot e^{-j \frac{2\pi \omega n}{N}}
\]

Where:
\begin{itemize}
    \item ${x}[m + n]$ is the input signal.
    \item $m$ is the time frame index.
    \item $n$ is the sample index within each frame.
    \item $w[n]$ is the window function.
    \item $\omega$ represents the frequency index.
    \item $N$ is the frame length.
\end{itemize}

\noindent \textbf{3. Feature Extraction:}

\begin{itemize}
    \item \textbf{MFCCs Calculation:} \\
    The MFCCs capture the short-term power spectrum of the signal using a series of triangular filters along the mel scale, which approximates the human auditory perception. The steps are:
    \begin{enumerate}
        \item Compute the power spectrum $P = \frac{1}{N} \left| X[k] \right|^2$, where $X[k]$ is the Discrete Fourier Transform (DFT) of the signal.
        \item Apply a mel filterbank to convert the power spectrum into the mel scale.
        \item Take the logarithm of the mel-scaled power spectrum.
        \item Apply the Discrete Cosine Transform (DCT) to generate MFCC coefficients.
    \end{enumerate}
    
The figure below (Figure~\ref{fig:mfcc_plot}) illustrates the Mel Frequency Cepstral Coefficients (MFCCs) extracted from the audio file:

\begin{figure}[H]
    \centering
    \includegraphics[width=0.7\textwidth]{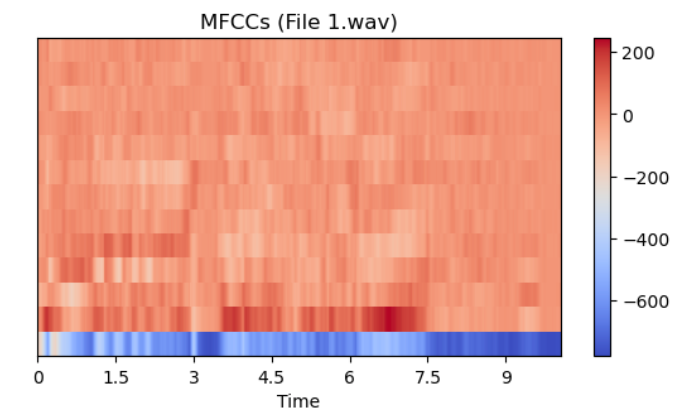} 
    \caption{MFCCs visualization for the audio file. The horizontal axis represents time (seconds), and the vertical axis represents different MFCC coefficients. The color bar on the right shows the intensity levels of the coefficients.}
    \label{fig:mfcc_plot}
\end{figure}

    \item \textbf{Melspectrogram Generation:} \\
    The Melspectrogram captures the distribution of energy over time and frequency. It is computed using the magnitude of the STFT coefficients:
    \[
    \text{Melspectrogram}(m, \omega) = \left| \text{STFT}\{ x[n] \}(m, \omega) \right|^2
    \]
\end{itemize}

The figure below (Figure~\ref{fig:melspec_plot}) illustrates the Mel-spectrogram for the audio file:

\begin{figure}[H]
    \centering
    \includegraphics[width=0.7\textwidth]{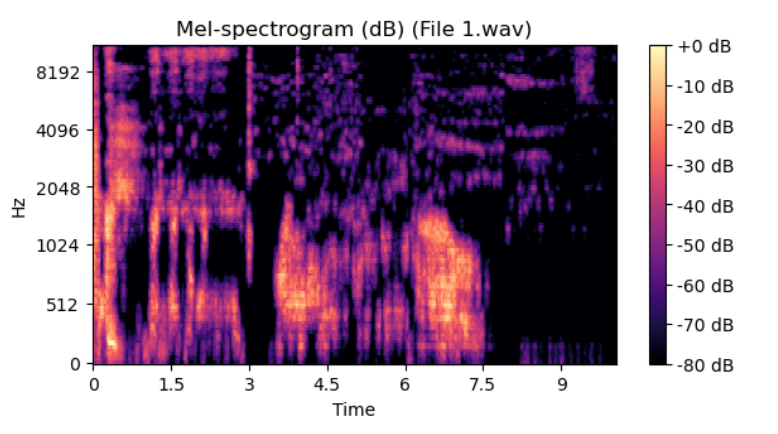} 
    \caption{Mel-spectrogram (dB) visualization for File 1.wav. The horizontal axis represents time (seconds), and the vertical axis represents frequency (Hz). The color bar on the right indicates the intensity in decibels (dB).}
    \label{fig:melspec_plot}
\end{figure}

\noindent \textbf{4. Feature Fusion:}

The MFCC coefficients were combined with the Melspectrogram matrix, creating a single, comprehensive feature vector for each frame. The combined feature set was then normalized using min-max scaling to ensure uniform scaling across all components.

\begin{figure}[H]
    \centering
    \includegraphics[width=0.7\textwidth]{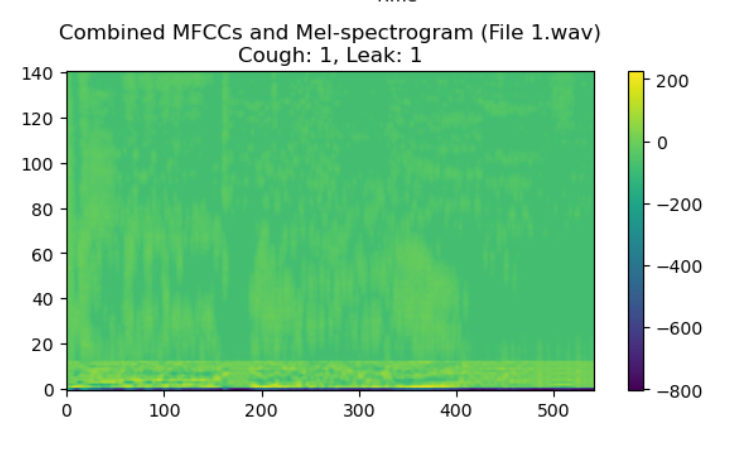} 
    \caption{Combined MFCCs and Mel-spectrogram visualization for File 1.wav. The title indicates detected events: Cough: 1, Leak: 1.}
    \label{fig:combined_mfcc_mel}
\end{figure}

The figure below (Figure~\ref{fig:combined_mfcc_mel}) shows the visualization of combined MFCCs and Mel-spectrogram for File 1.wav. The labels indicate the presence of coughs and leaks in the recording.

\subsubsection {Choosing a Model}
The goal was to select a model that could handle the complexity of the combined feature set and provide accurate classification results. Several models were considered and ANFIS was chosen for its ability to handle uncertainties and non-linear relationships \cite{kapoor2023prediction}. It integrates fuzzy logic with neural networks, using fuzzy if-then rules and membership functions to model complex relationships between the input features and target labels.

\subsubsection{Training and Evaluating the Model}
\begin{enumerate}[leftmargin=2em, label=\textbf{\arabic*.}]
    \item \textbf{Cross-Validation Setup:}
    \begin{itemize}
        \item The dataset was split into 5 folds for cross-validation. Each fold served as the validation set once, while the remaining four folds were used for training. This ensured that each data point was used for both training and validation, allowing for a comprehensive evaluation of the model's performance.
    \end{itemize}
    
    \item \textbf{Training the Model:}
    \begin{itemize}
        \item The ANFIS model was trained for 50 epochs per fold. During each epoch, the model parameters were updated using gradient descent to minimize the loss function:
    \end{itemize}
    
    \[
    \text{Loss} = \frac{1}{N} \sum_{i=1}^{N} (y_i - \hat{y}_i)^2
    \]
    
    Where:
    \begin{itemize}
        \item $y_i$ is the true label.
        \item $\hat{y}_i$ is the predicted label.
    \end{itemize}
    
    \item \textbf{Performance Metrics:}
    \begin{itemize}
        \item The primary evaluation metrics were \textbf{accuracy} and \textbf{loss} for both training and validation sets. Accuracy is defined as:
    \end{itemize}
    
    \[
    \text{Accuracy} = \frac{\text{Number of Correct Predictions}}{\text{Total Number of Predictions}}
    \]
    
    \item \textbf{Handling Overfitting:}
    \begin{itemize}
        \item Early stopping was monitored using the validation loss. If the loss did not improve for 10 consecutive epochs, training was halted early to prevent overfitting.
    \end{itemize}

    \item \textbf{Model Evaluation:}
    \begin{itemize}
        \item The final model was evaluated using the average performance metrics from all 5 folds. This included the mean accuracy, standard deviation, and overall loss.
    \end{itemize}
    
\end{enumerate}

\subsection{Method 2 - Spectral Analysis for Cough Detection}
\label{sec:method 2}
Following the segmentation of the denoised audio signal using the STFT, we generate a spectrogram that represents the signal’s time-frequency characteristics. The STFT is given by:

\[
\text{STFT}\{\tilde{x}[n]\}(m, k) = \sum_{n=0}^{N-1} \tilde{x}[m + n] \cdot w[n] \cdot e^{-j \frac{2\pi k n}{N}},
\]

where \( \tilde{x}[m + n] \) represents the denoised input signal, \( m \) is the time frame index, \( n \) is the sample index within each frame, \( w[n] \) is the window function, \( k \) is the discrete frequency index, and \( N \) is the frame length. This process generates a spectrogram, which is a time-frequency representation of the signal, allowing us to visualise the signal’s frequency content over time. 

Once the spectrogram is obtained, spectral analysis is applied to detect cough events in the audio. This detection process involves three main steps:

\begin{itemize}
    \item \textbf{Spectrogram Calculation:} 
    The spectrogram, which visually represents the frequency content of the signal over time, is created by calculating the squared magnitude of the STFT:

    \[
    \text{Spectrogram}(m, k) = |\text{STFT}\{\tilde{x}[n]\}(m, k)|^2.
    \]

    The squared magnitude of the STFT at each time frame \( m \) and frequency bin \( k \) indicates the signal’s energy at a specific time and frequency, enabling detailed analysis of the signal's temporal and spectral features.

    \item \textbf{Energy Computation:} 
    The energy of a discrete-time signal \( x(n) \), defined over a range \( N_1 \leq n \leq N_2 \), is given by \cite{chaparro2019discrete}:

    \[
    E_x = \sum_{n=N_1}^{N_2} |x(n)|^2.
    \]

    In the context of a spectrogram, the energy at each time frame is computed by summing the squared magnitudes of the STFT coefficients. These coefficients represent the signal’s amplitude and phase in the frequency domain. Therefore, the energy \( E[m] \) at time frame \( m \) is given by:

    \[
    E[m] = \sum_{k} |\text{STFT}\{\tilde{x}[n]\}(m, k)|^2,
    \]

    where \( \text{STFT}\{\tilde{x}[n]\}(m, k) \) are the STFT coefficients, representing the signal’s frequency content at time frame \( m \) and frequency bin \( k \). This energy measure is dimensionless, as it is derived from the squared magnitudes of the normalised signal.

    \item \textbf{Energy Thresholding:} 
    To detect potential cough events, a robust threshold is applied to distinguish these events from background noise. The threshold is defined as:

    \[
    \text{Threshold} = \mu_E + 3\sigma_E,
    \]

    where \( \mu_E \) represents the mean energy and \( \sigma_E \) is the standard deviation of the energy.
\end{itemize}

Finally, the top five time frames with the highest energy values are selected as potential cough events, and their corresponding times are recorded. These times are then compared against the energy threshold to determine whether they likely represent actual cough events.

\subsection{Method 3 - Machine Learning}
\label{sec:method 3}
\begin{figure}[!htb]
    \centering
    \includegraphics[scale = 0.35]{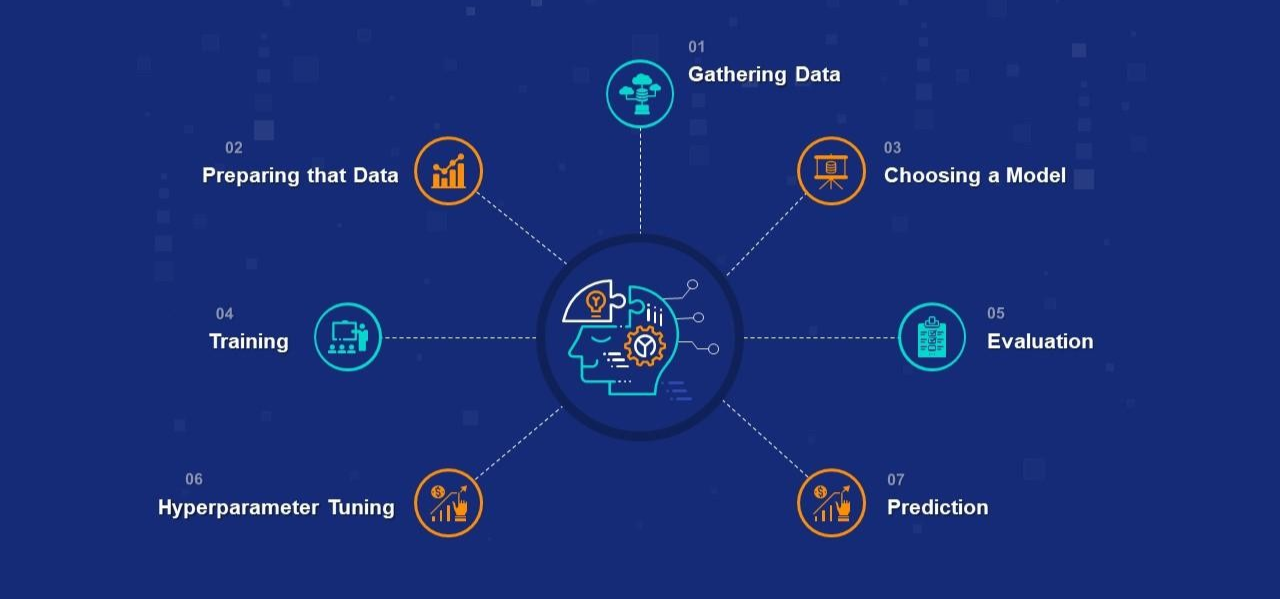}
    \caption{Machine Learning lifecycle to detect cough signals from audio data}
    \label{fig: ML lifecycle}
\end{figure}

The Machine Learning (ML) life-cycle followed in this study is illustrated in \ref{fig: ML lifecycle} and described in subsequent sections. The Scikit-Learn framework \cite{scikit-learn} has been used since it implements ML algorithms efficiently. 

\subsubsection {Gathering Data}
The audio signals (numbering 80) were provided by the manufacturer, labelled with `cough / no cough' and `leak \ no leak' against each signal. The signals were listened to for evaluating the audio quality and evaluating the need for audio pre-processing. It was concluded that in most cases, the cough is generally distinguishable; however, the leak was not conclusive. The cough signals are shown in Figure \ref{E70}.
\subsubsection{Preparing the data}
The data needs to be made suitable for Machine Learning to extract features so that the model is trained on them suitably. To analyse the audio signal, the librosa \cite{librosa} python package designed for audio analysis was used. It provides a rich functionality, whereby the audio signal can be analysed using time-series operations, spectrograms, time and frequency conversion, and pitch
operations. 

\begin{figure}[!htb]
    \centering
    \includegraphics[scale = 0.50]{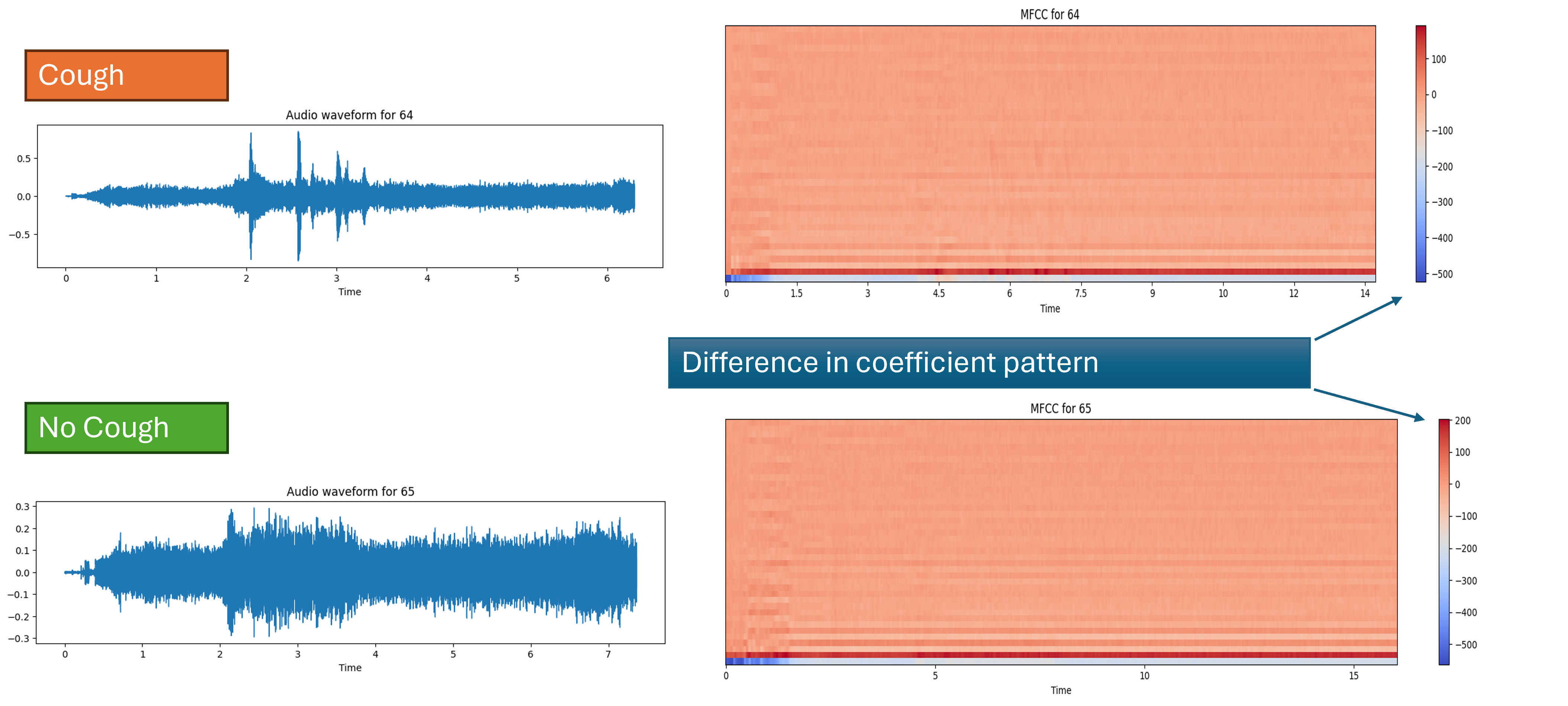}
    \caption{Mel-frequency Cepstral Coefficients (MFCCs) of the one of the 'cough' and 'no cough' audio signal showing distinct differences in the coefficient pattern}
    \label{fig: MFCC pattern}
\end{figure}

Based on research, there are a few representations that can detect coughs from spectral analysis of the audio signal. These are: Mel-frequency Cepstral Coefficients (MFCCs), Spectral Centroids (SC) and  Zero-Crossing Rate (ZRC). Several studies have demonstrated that MFCCs provide the highest correlation in detecting distinct coughs \cite{MFCCCough}\cite{alsabek2020studying}. 
In sound processing, the mel-frequency cepstrum (MFC) is a representation of the short-term power spectrum of a sound, based on a linear cosine transform of a log power spectrum on a nonlinear mel scale of frequency. MFCCs are coefficients that collectively make up an MFC and are derived from a type of cepstral representation of the audio clip (a nonlinear "spectrum-of-a-spectrum"). The difference between the MFC and the MFCC is that in the MFC, the frequency bands are equally spaced on the mel scale, which approximates the human auditory system's response more closely than the linearly-spaced frequency bands used in the normal spectrum. As shown in Figure \ref{fig: MFCC pattern}, there is a distinct difference in the pattern of MFCC for sample `cough' and `no cough' labelled signals.

\begin{figure}[!htb]
    \centering
    \includegraphics[scale = 0.45]{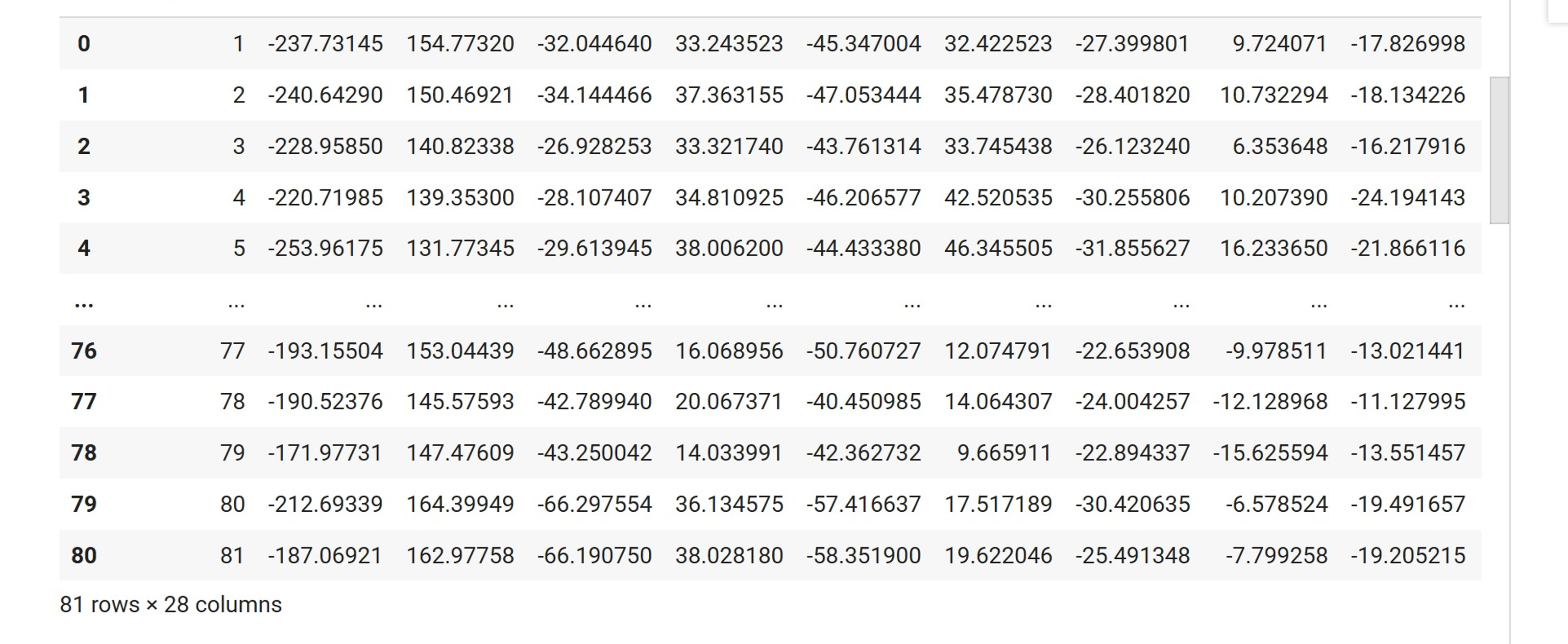}
    \caption{Feature vector generated from the MFCC of the cough audio signal}
    \label{fig: Feature}
\end{figure}

The next step is to extract the features of the MFCC using a combination of means and standard deviations of each frame, thus creating 26 attributes for each signal. Therefore the feature vector is an array of 81 (signals) x 26 (attributes), an extract of which is shown in Figure \ref{fig: Feature}.

\subsubsection{Choosing a Model}
Since the data is labelled and we are detecting a cough from the signal, a Supervised Learning Classification Model is suitable for this study. There are several models within this realm, and after reviewing various studies, the Support Vector Machine (SVM), Gradient Boost (XGBoost) and ANFIS: Adaptive Neuro-Fuzzy Network are found to be the most suitable \cite{balan2023explainable}. XGBoost is a decision-tree-based ensemble technique that minimizes the models' residuals and increases the predictive power by combining the predictions of several base estimators built with a given learning algorithm in order to improve generalizability / robustness over a single estimator. These factors have made this model more efficient than other machine learning models for their prediction accuracy in audio signals.

\subsubsection{Training and Evaluating the model}
\begin{figure}[!htb]
    \centering
    \includegraphics[scale = 0.8]{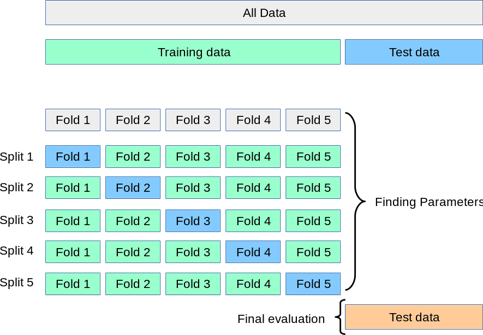}
    \caption{Cross-validation, where the training set is randomly split into 20 sub-sets and the model is trained on each of these sub-sets, resulting in better generalization.}
    \label{fig: Cross-val}
\end{figure}

Training a prediction function and evaluating it on the same dataset is a methodological error: a model that merely memorizes the labels of the samples it has been trained on would achieve a perfect score but would fail to generalize to new, unseen data. This issue is known as over-fitting. To prevent over-fitting, it is standard practice in supervised machine learning experiments to reserve a portion of the data as a separate test set, denoted as $X_test$ and $y_test$, for evaluation purposes. Therefore, the data-set  is split in the ratio of 80:20 using the $train_test_split$ function divide into a training set and a test set. Furthermore, Scikit-Learn`s K-fold cross-validation feature \ref{fig: Cross-val} was used to randomly splits the training set into 20 distinct subsets, or folds, and then trains and evaluates the model 20 times, each time using a different fold for evaluation and the remaining 19 folds for training. This process results in an array of 20 evaluation scores. Cross-validation not only provides an estimate of your model`s performance but also offers a measure of the estimate`s precision, such as its standard deviation.

Default parameters of the model were used, and no hyper-parameter tuning was required.

\subsubsection{Prediction}
The model was then used to generate predictions on the independent test-set, and the accuracy, precision results were generated against each iteration along with their standard deviation.

\subsection{Method 4 - Wavelet denoising}
\label{sec:spectral-denoising}
\subsubsection{Spectrogram Analysis}
\begin{figure}[ht]
    \centering
    \includegraphics[width=0.6\textwidth]{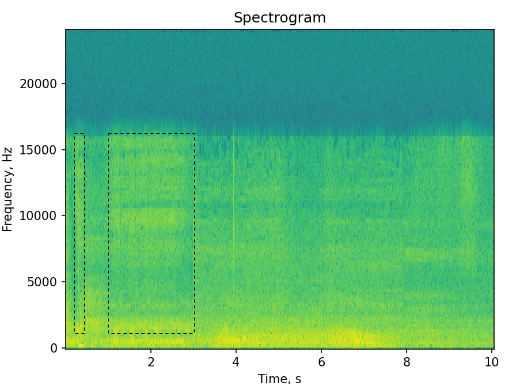}
    \caption{Spectrogram of a recording containing a cough. The regions corresponding to a pressure change and subsequent coughs are marked by dashed boxes. }
    \label{fig:spectrogram}
\end{figure}

Pressure changes and coughs are audible in the recordings where they are present, despite significant background noise. To visualise these we plotted the spectrogram of the recordings, which displays the frequency content of the recording over time as defined in \ref{sec:method 2}. An example recording is shown in Fig \ref{fig:spectrogram}. 

The pressure changes and coughs are seen as vertical bands in the spectrogram, due to their transient nature and full-spectrum frequency content. However, the signal-to-noise ratio is relatively poor, so further processing is required to separate these signals from the background. 

\subsubsection{Denoising}
A noisy signal can be written as the sum of the smooth underlying signal and a stochastic noise term: 
$$
S_{noisy}(t) = S_{smooth}(t) + \sigma.
$$

\begin{figure}[ht]
    \centering
    \includegraphics[width=0.55\textwidth]{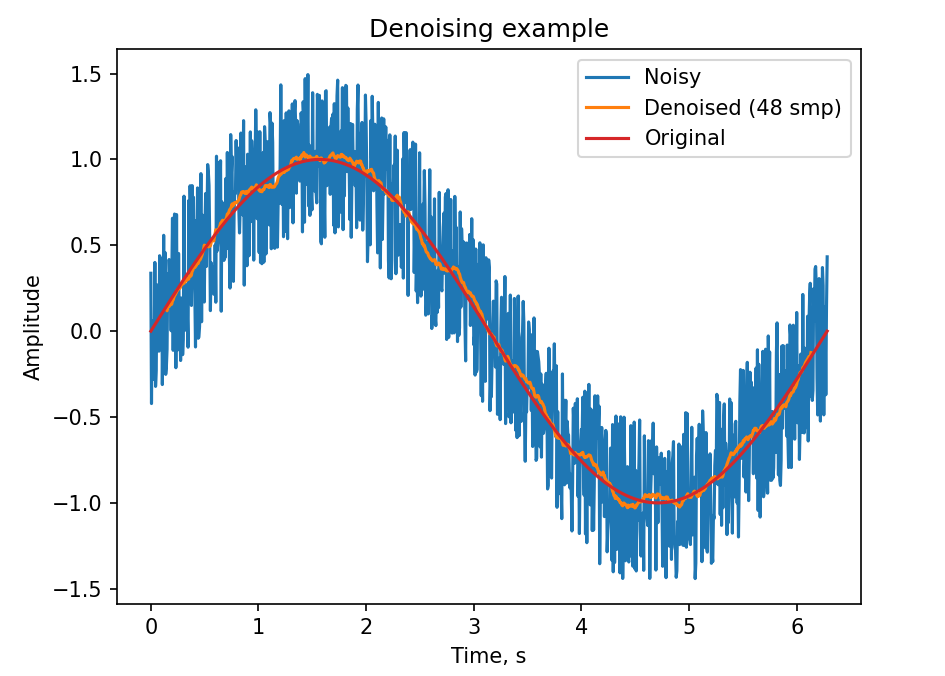}
    \caption{Example showing how a function can be denoised. Our original signal, a sine wave, is modulated by a noise term. Taking the moving average of the noisy signal restores the original. Using more samples gives a better approximation to the original signal. }
    \label{fig:denoise}
\end{figure}

Standard denoising techniques rely on smoothing the noisy signal by calculating a moving (possibly weighted) average. The assumption is that the noise term takes a random value with each sample, so that over a sufficiently large window, the contribution of the noise averages to 0, as shown in Fig \ref{fig:denoise}.   

\begin{figure}[ht]
    \centering
    \includegraphics[width=\textwidth]{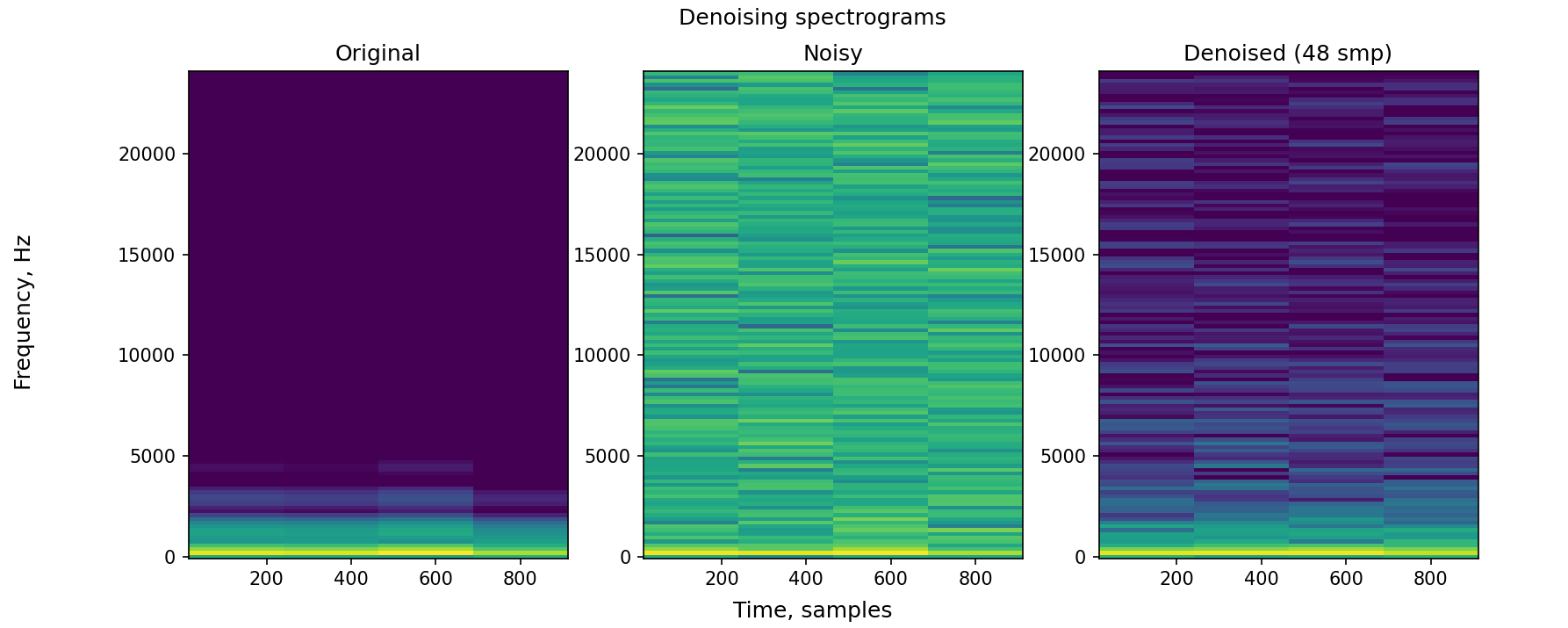}
    \caption{Left: the STFT of a pure sine wave is just a single line at the original frequency (plus some aliasing artifacts). Centre: pure white noise has a flat spectrum, i.e. a constant amplitude over all frequencies. Right: denoising preferentially removes higher frequencies.}
    \label{fig:denoise2}  
\end{figure}

Such smoothing methods work by filtering high-frequency information from the signal, as shown in Fig \ref{fig:denoise2}. They are therefore most effective when there is good separation between the spectra of the signal and the noise, and unsuitable when the signal and noise occcupy the same frequency range, as the signal would also be smoothed over. Unfortunately, as the spectrogram in Fig \ref{fig:spectrogram} shows, in this case the signal and noise spectra do overlap, so alternative denoising methods must be used.

\paragraph{Spectral gating}

One approach is to use spectral gating, as developed by Sainburg et al. \cite{spectral_gating}. We used their Python package \texttt{noisereduce} to denoise the recordings \cite{tim_sainburg_2019_3243139}. This algorithm takes two inputs: a signal clip, and a noise clip containing only the background noise. We used the signal clip for both inputs, as this was stated to work well, and meant that noise clips did not have to be isolated from each recording.
The algorithm works as follows:
\begin{enumerate}
    \item Calculate the spectrograms $S_s$ and $S_n$ of the signal and noise audio clips respectively.
    \item Caclulate the mean and standard deviation of each frequency component of $S_s$ and $S_n$ over time.
    \item Compute the short-time Fourier transform (STFT) of $S_n$.
    \item Compute a threshold noise level for each frequency component of $S_n$, based on the mean and standard deviation previously calculated.
    \item Generate a mask over $S_s$, using the thresholds determined from $S_n$.
    \item Smooth the mask over frequency and time, then apply to $S_s$ to remove noise.
    \item Compute the inverse STFT of $S_s$, to give the denoised signal.
\end{enumerate}

\begin{figure}[ht]
    \centering
    \includegraphics[width=\textwidth]{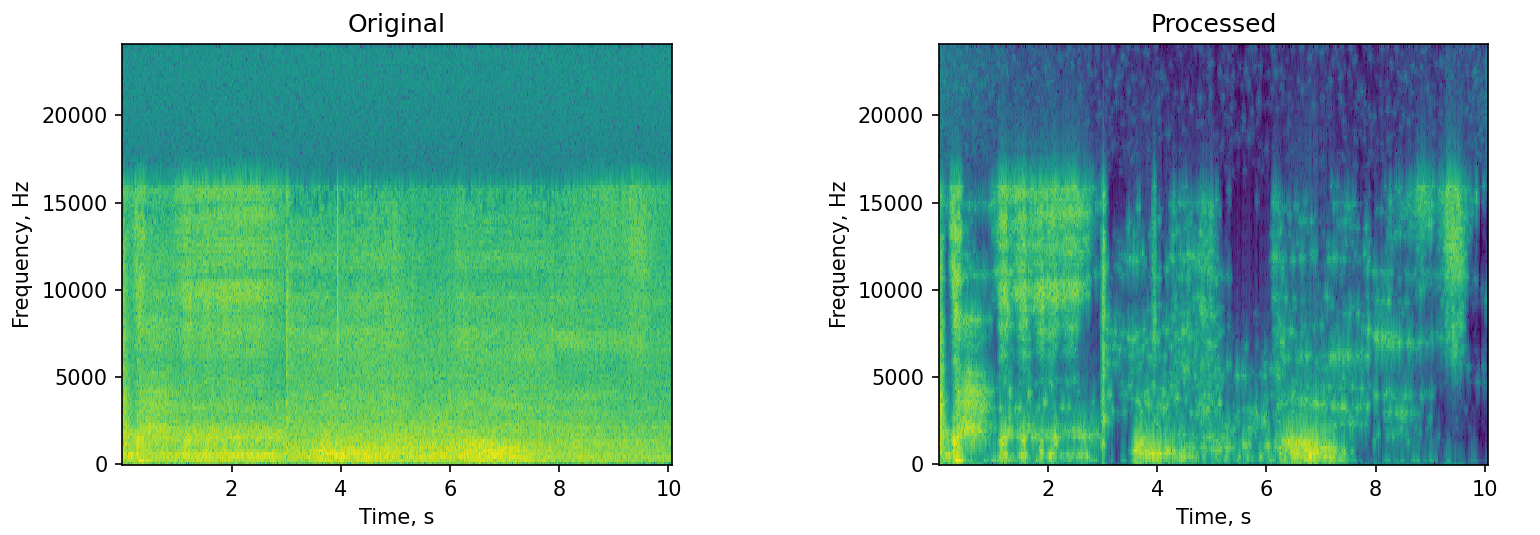}
    \caption{Results of applying the \texttt{noisereduce} algorithm to a recording.}
    \label{fig:spectral_gating}  
\end{figure}

The overall effect is to apply a noise gate (with a time-varying threshold based on the magnitude of the noise), to each frequency component of the signal clip. This results in a modest amount of denoising, but this was insufficient to fully suppress the background noise, as shown in Fig \ref{fig:spectral_gating}. Better results may be achieved by isolating a noise clip from each recording, or tuning other parameters used by the \texttt{noisereduce} algorithm.

\paragraph{Spectral subtraction}

We also used spectral subtraction to more aggressively denoise the recordings, at the cost of potentially removing some of the desired signal. We can write the signal $S(t)$ as follows:

$$ 
S(t) = P(t) + C(t) + L(t) + \sigma 
$$

where $P(t)$, $C(t)$ and $L(t)$ are the signals from pressure changes, coughs and leaks respectively, and $\sigma$ is the machine noise term. Taking the Fourier transform gives:

$$ 
\hat{S}(\omega) = \hat{P}(\omega) + \hat{C}(\omega) + \hat{L}(\omega) + \hat{\sigma}
$$

\begin{figure}[ht]
    \centering
    \includegraphics[width=\textwidth]{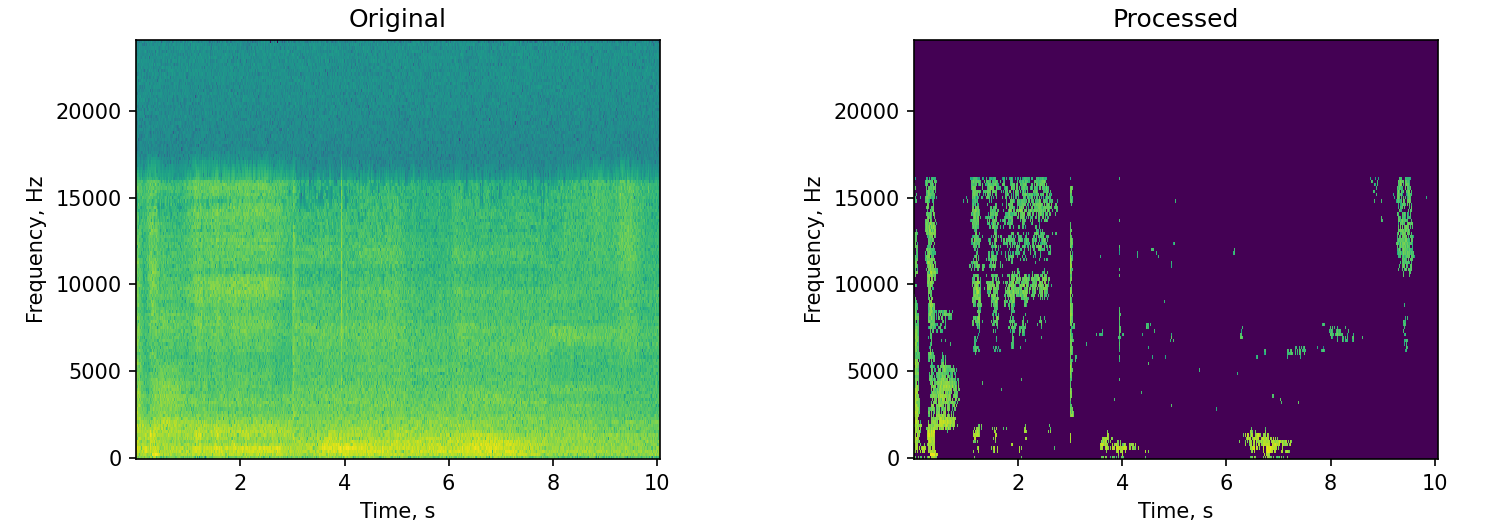}
    \captionof{figure}{Results of applying the spectral subtraction algorithm to a recording.}
    \label{fig:spectral_subtraction}  
\end{figure}

In the recordings, pressure changes and coughs have high amplitude, but are transient signals that only occur over short times. Leaks are the opposite: sustained signals with low amplitude. The only term which has high amplitude throughout the recording is the noise term. Therefore, given a sufficiently long recording, the contribution of the noise term dominates. Assuming that the noise term is static (that its spectrum is constant over the recording), we can approximate the noise spectrum by calculating the mean amplitude of the signal for each frequency component of the STFT. Subtracting this from the signal spectrum at each time point, then taking the inverse STFT gives the desired denoised signal, as shown in Fig \ref{fig:spectral_subtraction}. This is essentially equivalent to the \texttt{noisereduce} algorithm, but using a fixed noise gate with the threshold set at the mean of each frequency component. 

While this does remove much more machine noise, the contribution to the signal from sustained sounds such as leaks, are also removed. The spectral gating algorithm removes too little background, while this method removes too much signal. The same techniques could be applied to either method to achieve a suitable balance of noise removal to signal preservation, such as isolating snippets of just machine noise, so that we can more accurately calculate and remove its spectrum.  

\subsubsection{Cough detection}
With the recordings adequately denoised, we attempted to detect coughs in the residual signal. Coughs and pressure changes are transient sounds, so we applied a modified version of the transient detection algorithm of Thoshkahna et al. \cite{transient_detect}. This modified algorithm is as follows:
\begin{enumerate}
    \item Compute the absolute value of the STFT of the signal. Fortunately, this is trivial to calculate given the spectrogram, as discussed in \ref{sec:method 2}
    \item Compute $\sigma_i$ by summing over the frequency axis of the STFT:
        $$ \sigma_i \equiv \sigma(t_i) = \sum_j |\textrm{STFT}_{ij}|,$$
        where $\textrm{STFT}_{ij}$ is the $j$th frequency component of the STFT at time $t_i$.
    \item Calculate functions $T^+_i$ and $T^-_i$, where
        $$T^+_i = \sigma_i -\sigma_{i+1}$$ 
        $$T^-_i = \sigma_i -\sigma_{i-1}.$$
    \item Calculate $F_i$, which is given by
        $$F_i = \frac{1}{2} (T^+_i(1+\textrm{sgn}(T^+_i)) + T^-_i(1+\textrm{sgn}(T^-_i))),$$
        where $\textrm{sgn}(x)$ is the usual sign function.
    \item Calculate an adaptive threshold $\lambda$:
        $$\lambda_i = \beta \left( \frac{1}{N}\sum_{k=i-\frac{N}{2}}^{i+\frac{N}{2}} F_j \right),$$
        where $\beta$ is a parameter that sets the height of the threshold, and the bracketed term is simply the moving average of $F$ over a window of width $N$. We used $\beta=0.5$ and $N=10$. 
    \item Define transient regions where $F_i > \lambda_i$.
\end{enumerate}

The main difference between this algorithm and the one described in \cite{transient_detect} is that we sum over the whole frequency spectrum before calculating the gradients $T^+$ and $T^-$, since we only need to locate full-spectrum transients from coughs and pressure changes, rather than extract transients from the original signal.

\begin{figure}[ht]
    \centering
    \includegraphics[width=0.6\textwidth]{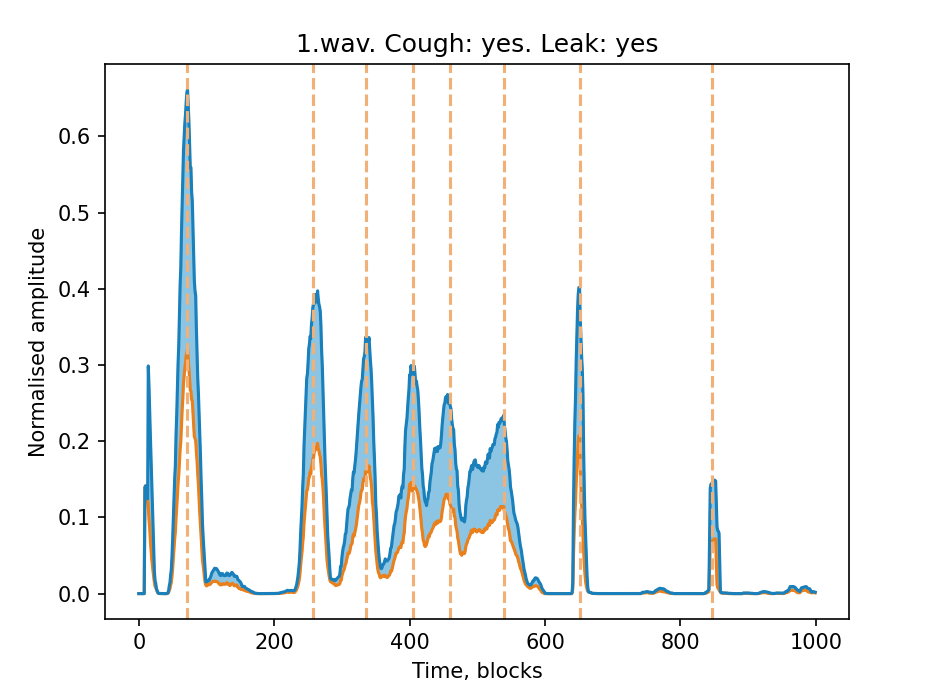}
    \caption{Example of the peak detection algorithm applied to a recording.}
    \label{fig:peak_detect}  
\end{figure}

An example of this transient detection is shown in Fig \ref{fig:peak_detect}. We located transient peaks using the \texttt{find\_peaks} function in the \texttt{scipy.signal} Python package \cite{scipy}, with the peak width, height and prominence set to 30, 0.02 and 0.01 respectively. Once transients were located, the recordings could be classified into ``cough'' and ``no cough'' categories. We had hoped to combine the approach taken in this section with the machine learning work presented in section \ref{sec:method 3}, by using the locations of detected transients as inputs to a neural-net based classifier. As time was limited, however, we implemented a rudimentary classification based on the following principles:
\begin{itemize}
    \item A cough is most likely to occur immediately following a pressure change from the MI-E device.
    \item Coughs often consist of a double sound, where two explosive phases occur in a short period ($<1$s) \cite{KORPAS1996261,Kelsall175}.
\end{itemize}
We therefore classified recordings as containing a cough if they contained 3 transients occurring in a 2s window, representing a double cough sound following a machine pressure change.

\subsection{Method 5 - Predictive modelling}
\label{sec:model}

We consider a mathematical model of the flow of air through the lung in response to a prescribed inspiration pressure which mimicks that imposed by a mechanical insufflation device. Our aim is to predict the corresponding flow rate through this device, from which we can then assess the response of the system to leaks.

\begin{figure}[H] 
\centering
\includegraphics[width=\textwidth]{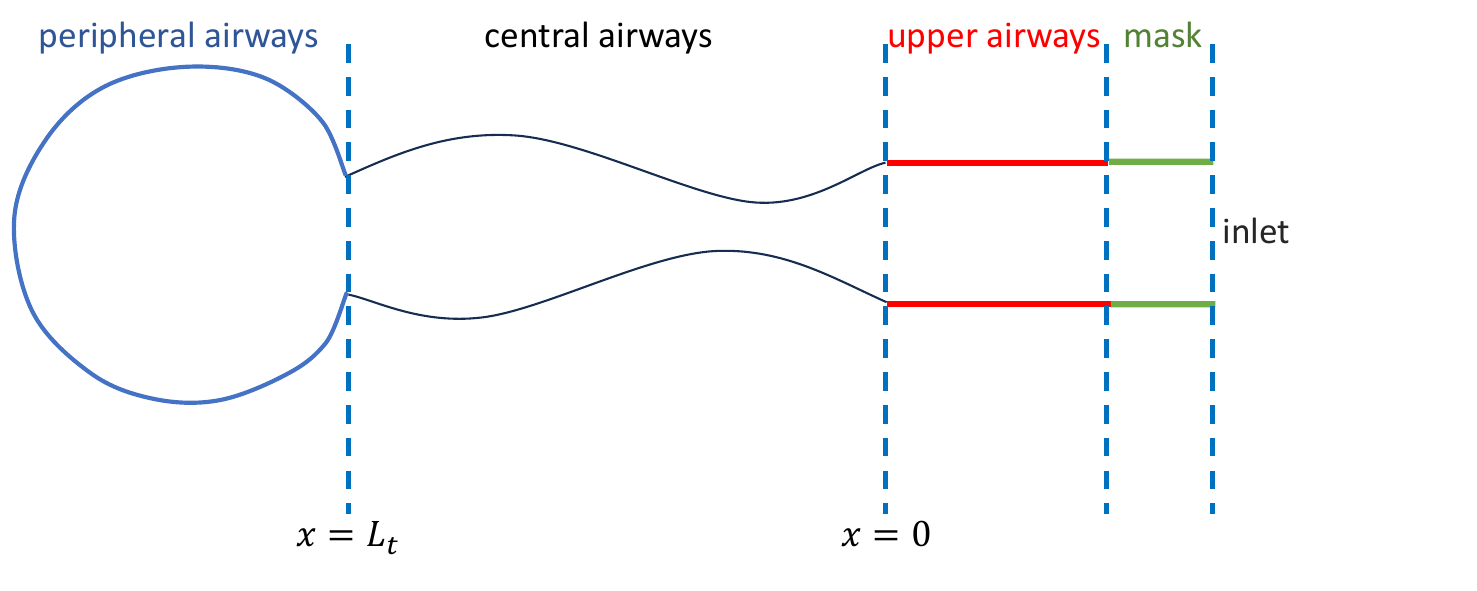} 
\caption{Sketch of the model setup.}
\label{fig:sketch}
\end{figure}

Our approach follows a previous model of the dynamic cough maneuvers in the human lung \cite{naire2009dynamics}; we divide the lung into four segments arranged in series, as sketched in Fig.~\ref{fig:sketch}. Adjacent to the inlet we consider a region which mimics the mask of the insufflation device (where leakiness may subsequently arise), modelled as a rigid cylindrical tube of radius $R^\ast_c$ and length $L^\ast_m$; the pressure supplied by the insufflation device is prescribed as a boundary condition on the outer circular surface of this mask region. The other end of the mask region connects into the upper airways, which are again modelled as a rigid cylindrical tube of length $L^\ast_u$ and radius $R^\ast_u$. This region adjoins a flexible-walled tube of baseline radius $R^\ast_c$ and length $L^\ast_c$ which models the central airways. This tube terminates in a closed ended region formed by a compliant bag which mimics the peripheral airways.

Our attention focuses primarily on the flow of air along the central airway region \cite{naire2009dynamics}. We assume that this tube is a uniform circular cylinder in its baseline state with cross-sectional area $A^\ast_c = \pi {R^\ast_c}^2$, parameterised by the coordinate $x^\ast$ oriented along the tube mid-line. We denote $x^\ast=0$ at the inlet of central airways

We assume that the vessel is subject to a spatially uniform external pressure $P_e^\ast$. As the cross-section of this vessel remains circular as it deforms, with radius $r^\ast=R^\ast(x^\ast,t^\ast)$, where $t^\ast$ is time; the corresponding cross-sectional area of the vessel can be written $A^\ast(x^\ast,t^\ast) = \pi {R^\ast}^2$. 

We assume the flow of air has uniform density $\rho^\ast$ and is nearly inviscid, approximated by a uniform plug flow profile $U^\ast(x^\ast,t^\ast)$. The corresponding air pressure is denoted $P^\ast(x^\ast,t^\ast)$.

We assume this central airway region is long compared to its baseline radius, which results in a significant reduction in the flow equations. Following Naire \cite{naire2009dynamics}, we express the governing equations across the central airway region ($0\le x^\ast \le L_c^\ast$) in the form
\begin{subequations}
\label{eq:gov}
\begin{align}
A^\ast_{t^\ast} + (U^\ast A^\ast)_{x^\ast} &=0,\\
\rho^\ast \left(U^\ast_{t^\ast} + U^\ast U^\ast_{x^\ast}\right) & = -P^\ast_{x^\ast} + F^\ast(A^\ast,U^\ast),
\end{align}
where the function $F^\ast$ incorporates the (small) viscous resistance of the flow. In this paper we consider the simple form
\begin{equation}
F^\ast(A^\ast,U^\ast) = -k \mu^\ast \frac{U^\ast}{A^\ast}.
\end{equation}
where $k$ is a viscous resistance parameter and $\mu^\ast$ is the fluid viscosity.

Within the tube wall we incorporate an elastic resistance to deformation; following Naire \cite{naire2009dynamics}, we model the tube wall with stiffness $E^\ast$ and longitudinal tension $T^\ast$, in the form
\begin{equation}
P^\ast = P_e^\ast + E^\ast {\cal P}(A^\ast/A^\ast_c) - T^\ast A^\ast_{x^\ast x^\ast},
\end{equation}
where we assume the functional form (or `tube law')
\begin{equation}
{\cal P}(s) = s^m - s^{-n},
\end{equation}
where $m,n \ge 0$ are non-negative exponents.
\end{subequations}

In cases where the mask is not leaking then the mask and upper airway regions are essentially identical and can be conflated into one region ($-L^\ast_m - L_u^\ast \le x^\ast \le 0$), where the cross-sectional area of the conduit is uniform. In these regions we apply the governing equations (\ref{eq:gov}), where for rigid walls the corresponding flow $U^\ast$ is spatially uniform and the equations reduce to
\begin{subequations}
\label{eq:bcs}
\begin{align}
\label{eq:gov_rigid}
\rho^\ast U^\ast_{t^\ast} & = -P^\ast_{x^\ast} + F^\ast(A_c^\ast,U^\ast).
\end{align}

At the inlet to the system we impose an inlet pressure  $P^\ast = P_{in}^\ast(t^\ast)$. This function is prescribed according to a typical pressure ramping protocol prescribed by a mechanical insufflation device, involving fixed periods of positive and negative pressure with short intervals of zero pressure in between. A typical time-dependent profile is shown as the black solid lines in Fig.~\ref{fig:pressure_time}.

To close the system of equations we impose that the area of the central airway segment must match the area of the rigid segment at their junction, so that
\begin{equation}
A^\ast=A_c^\ast, \qquad (x^\ast=0).
\end{equation}
Further applying continuity of flow rate and pressure between the central airway region and the upper airway region, we hence integrate (\ref{eq:gov_rigid}) to compute a boundary condition on the inlet to the central airway region in the form
\begin{equation}
T^\ast A^\ast_{x^\ast x^\ast} = P_e^\ast - P_{in}^\ast(t^\ast) + (L^\ast_m + L^\ast_u)(\rho^\ast U^\ast_{t^\ast} - F^\ast(A_c^\ast,U^\ast)), \qquad (x^\ast=0).
\end{equation}
Downstream of the central airways we model the peripheral airways as a compliant bag of volume $V^\ast_p(t^\ast)$. For simplicity we impose that the area of the central airway segment is fixed at the junction with this bag, so that
\begin{equation}
A^\ast=A_c^\ast, \qquad (x^\ast=L^\ast_c).
\end{equation}
The flow of air moving from the central airways into the periphery must balance the rate of increase of volume of the periphery, so that
\begin{equation}
\label{eq:volode}
\frac{dV_p^\ast}{dt^\ast} = U^\ast(L^\ast_c,t^\ast) A_c^\ast.
\end{equation}
The corresponding pressure in the peripheral airways is determined from the volume of this region via a tube law, where we assume
\begin{equation}
P^\ast_p - P^\ast_{p0} = \frac{1}{C_p^\ast} (V^\ast_p - V^\ast_{p0}),
\end{equation}
where $P^\ast_{p0}$ is a reference peripheral airway pressure, $V^\ast_{p0}$ is the corresponding reference volume and $C_p^\ast$ is the compliance of the peripheral airways. Taking these reference values to be equal to their initial values (before any pressure perturbation is applied to the system), we integrate (\ref{eq:volode}) to obtain
\begin{equation}
V_p^\ast(t^\ast) = V^\ast_{p0} + A_c^\ast \int_0^{t^\ast} U^\ast(L^\ast_c,t) \,dt.
\end{equation}
Finally, we close the entire system by enforcing continuity of pressure between the outlet from the central airways and the compliant bag, so that
\begin{equation}
P^\ast(L^\ast_c,t^\ast) = P_p^\ast(t^\ast), \qquad (x^\ast=L^\ast_c).
\end{equation}
\end{subequations}
This system of equations is solved using a numerical method analogous to that presented by \cite{stewart2009local}.

Predictions of the model are shown in Fig.~\ref{fig:pressure_time} below, where we compute the traces of pressure at the inlet and outlet of the central airway segment (blue and red lines, respectively), as well as the flow rate of air supplied by the insufflation device (green line) and into the peripheral airways (magenta line). Note that this air flow rate is oriented into the lung, and so the aim of the maneuver is to generate a large negative $q_{in}(t)$, which would then transport mucus out of the lung.

The dimensional parameters for the model follow from Naire \cite{naire2009dynamics} and are listed in Table \ref{tab:paras}.

\begin{table}[h!]
\centering
\begin{tabular}{|c|c|c|}
\hline
Parameter & Symbol  & Value\\ 
\hline
Length of mask region  & $L^\ast_m$ & 0.05m\\
Length of upper airway region & $L^\ast_u$ & 0.05m \\
Length of central airway region & $L^\ast_c$ & 0.1m \\
Radius of central airway region & $A^\ast_c$ & 0.01m \\
Density of air & $\rho^\ast$ & 1.225 kg/m$^3$\\
Dynamic viscosity of air & $\mu^\ast$ & $3.178\times10^{-5}$ kg/m/s \\
Baseline volume of lung periphery & $V^\ast_{p0}$ & 2.5l \\
Baseline pressure of lung periphery & $P_{p0}^\ast$ & 0cmH$_2$O \\
External pressure on central airways & $P_e^\ast$ & 0cmH$_2$O \\
Stiffness of central airways & $E^\ast$ & $10^{4}$ Pa\\
Wall tension in central airways & $T^\ast$ & 10$^{5}$ Pa\\
Compliance of peripheral airways & $C^\ast_{p}$ & 0.1 l/cmH$_2$O \\
\hline
\end{tabular}
\caption{Parameter values for the predictive model}
\label{tab:paras}
\end{table}


\section{Results} 
\label{sec:results}

\subsection{Results 1 - Adaptive Neuro-Fuzzy Inference System (ANFIS)}
\label{sec:results 1}

\begin{figure}[H] 
    \centering
    \includegraphics[width=0.6\textwidth]{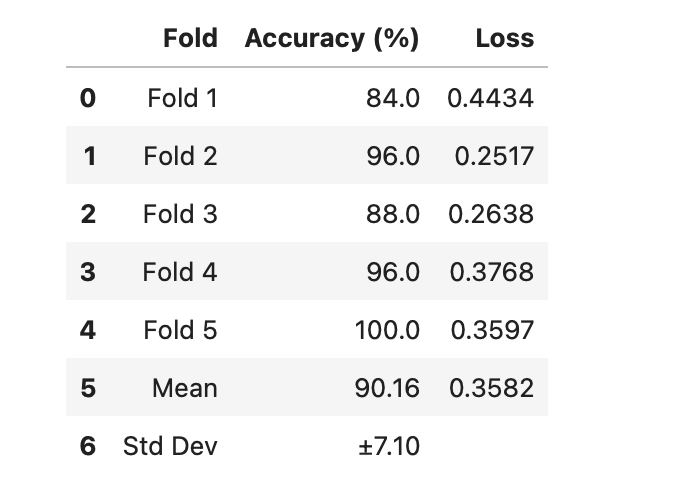} 
    \caption{Cross-validation results for the ANFIS model, showing accuracy and loss for each fold. The mean and standard deviation are calculated to provide an overall performance evaluation.}
    \label{fig:table_image}
\end{figure}

The image below (Figure~\ref{fig:table_image}) shows the cross-validation results for the ANFIS model. The table highlights the accuracy and loss for each fold, along with the mean and standard deviation values. Despite some fluctuations, the overall model performance was stable. Combining MFCCs and Melspectrograms into a single feature vector significantly improved classification performance for respiratory sounds. The ANFIS model demonstrated strong generalization capabilities and consistent performance across multiple folds.

\subsection{Results 2 - Spectral Analysis for Cough Detection}

\label{sec:results 2}

\begin{figure}[!htb]
\centering
    \makebox[\textwidth]{
        \includegraphics[width=\textwidth]{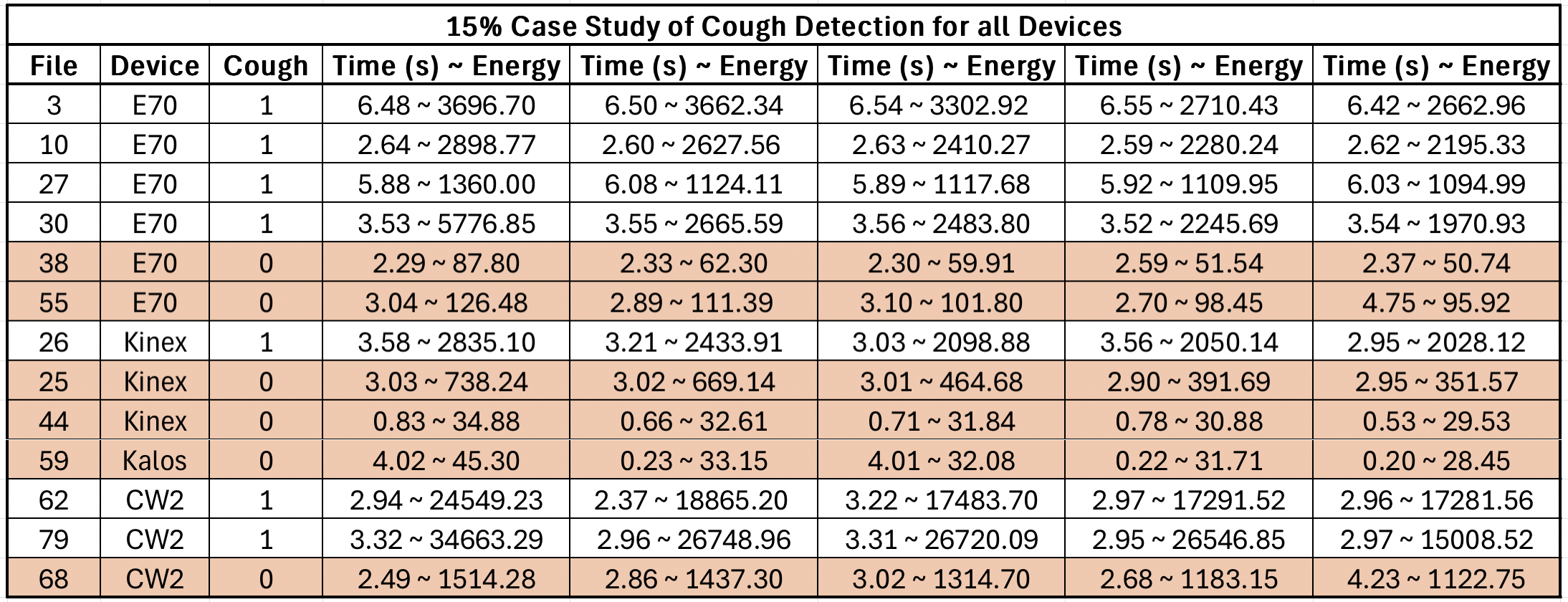}
    }
    \caption{Case study of cough detection across all devices, with false positives highlighted in orange.}
    \label{Cough CaseStudy}
\end{figure}

\autoref{Cough CaseStudy} presents a case study on cough detection, which covers 15\% of all devices. For each file analysed, it provides details on the timing of detected coughs and their corresponding energy levels, listing up to the top five highest energy levels. Although some detections occurred in instances where no cough was present (i.e., cough = 0), these represent false positive detections. This conclusion is drawn from the consistently low energy levels recorded across all devices in these instances, in contrast to the significantly higher energy levels associated with true positive cases (cough = 1). Therefore, proper comparisons should be made between scenarios such as E70 (cough) versus E70 (non-cough), rather than E70 versus CW2. It is also noted that, across all devices, CW2 consistently displayed the highest energy levels.

\begin{figure}[!htb]
\centering
\begin{subfigure}{.49\textwidth}
    \centering
    \includegraphics[width=\textwidth]{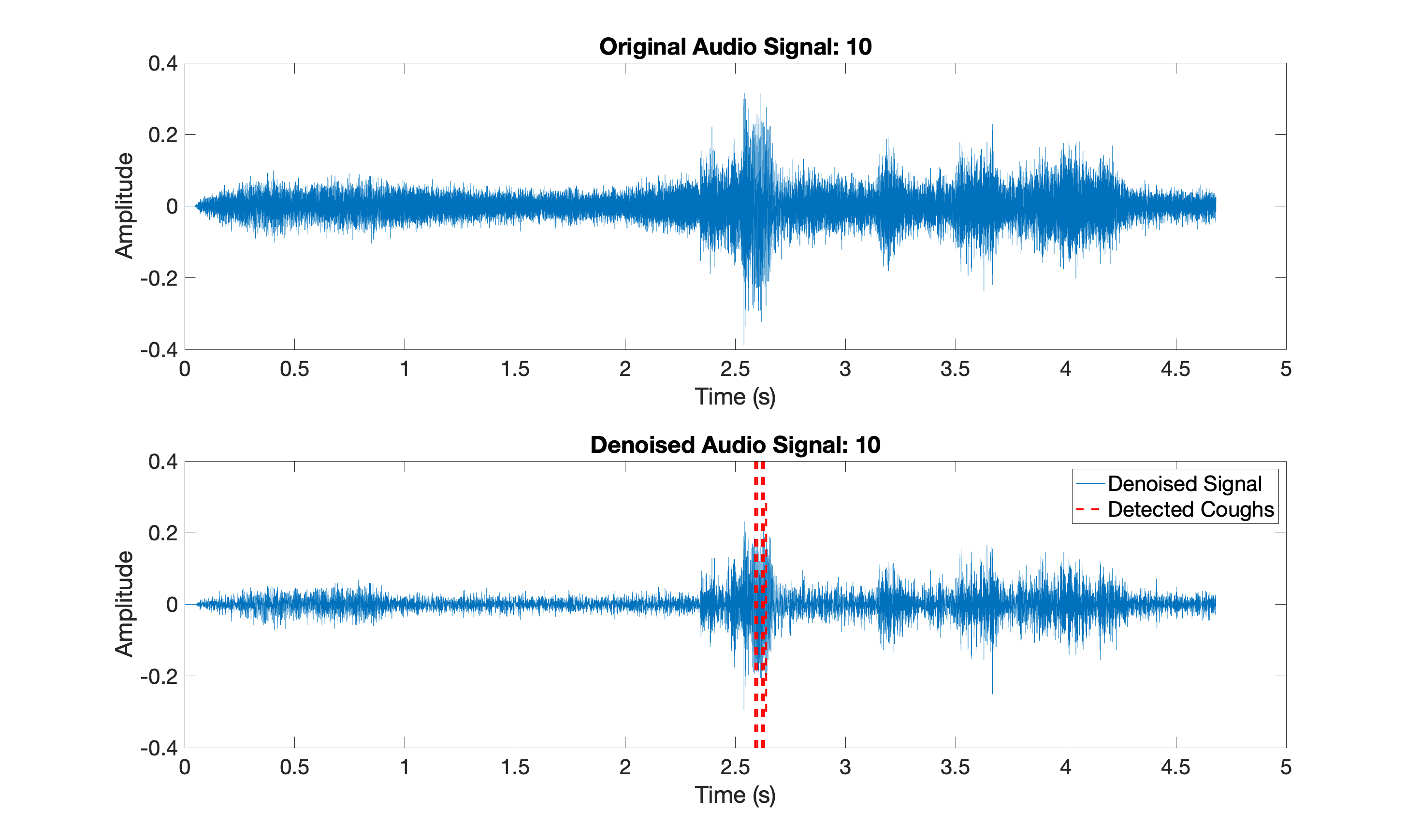}  
    \caption{Cough 1: True Positive}
    \label{Cough 10}
\end{subfigure}
\begin{subfigure}{.45\textwidth}
    \centering
    \includegraphics[width=\textwidth]{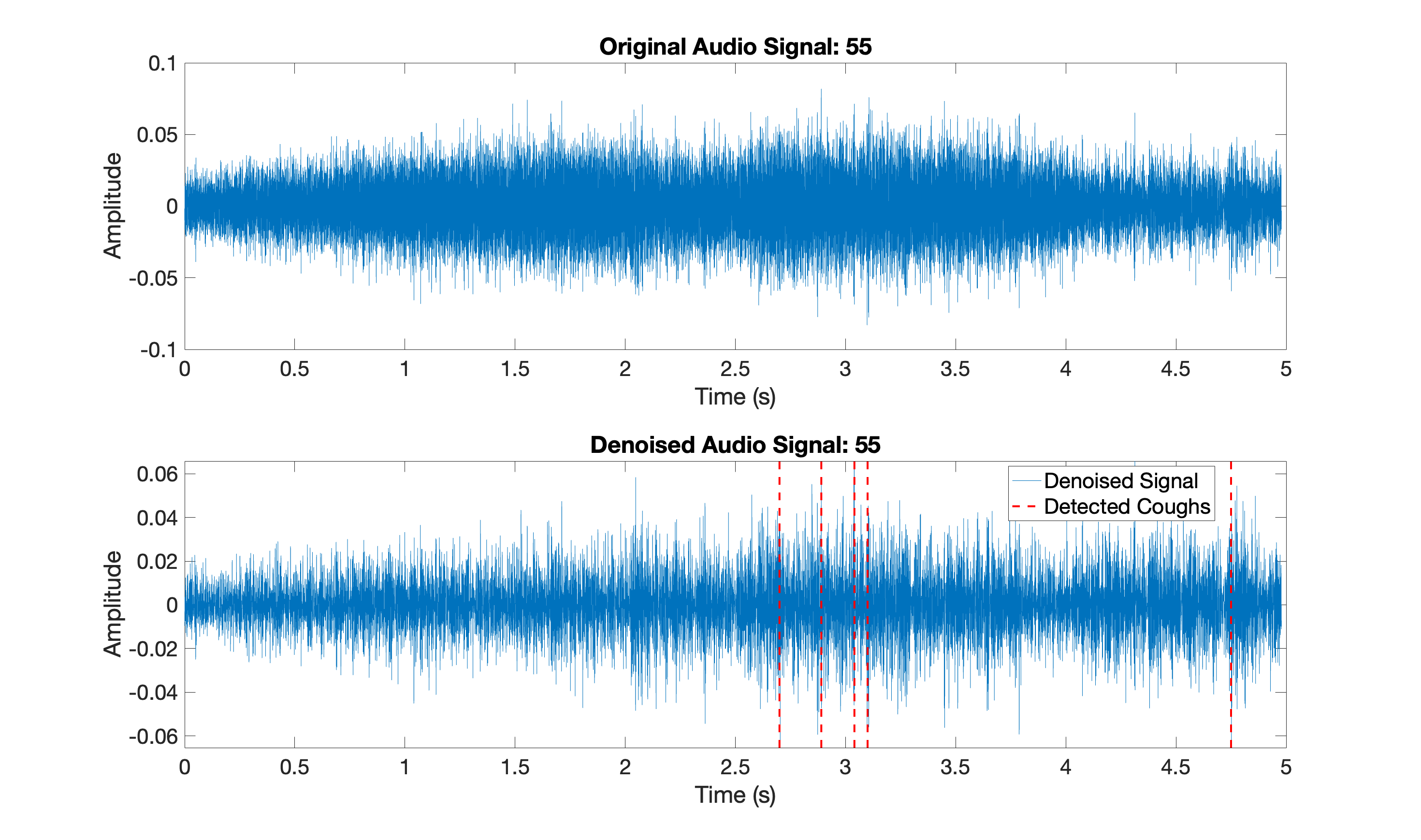}  
    \caption{Cough 0: False Positive}
    \label{No cough 55}
\end{subfigure}
\caption{Cough detection in E70 devices.}
\label{E70}
\end{figure}

\begin{figure}[!htb]
\centering
\begin{subfigure}{.49\textwidth}
    \centering
    \includegraphics[width=\textwidth]{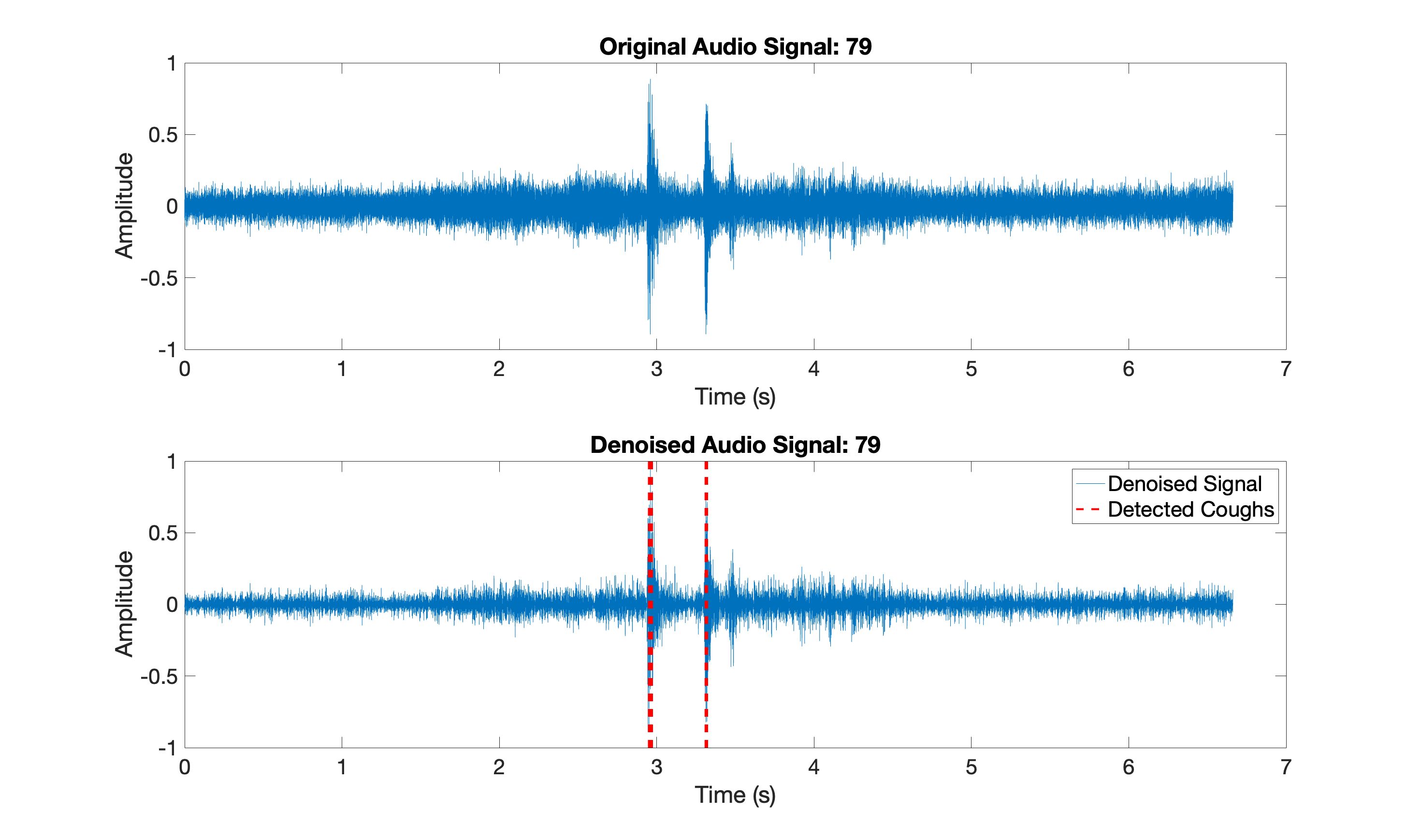}  
    \caption{Cough 1: True Positive}
    \label{Cough 79}
\end{subfigure}
\begin{subfigure}{.49\textwidth}
    \centering
    \includegraphics[width=\textwidth]{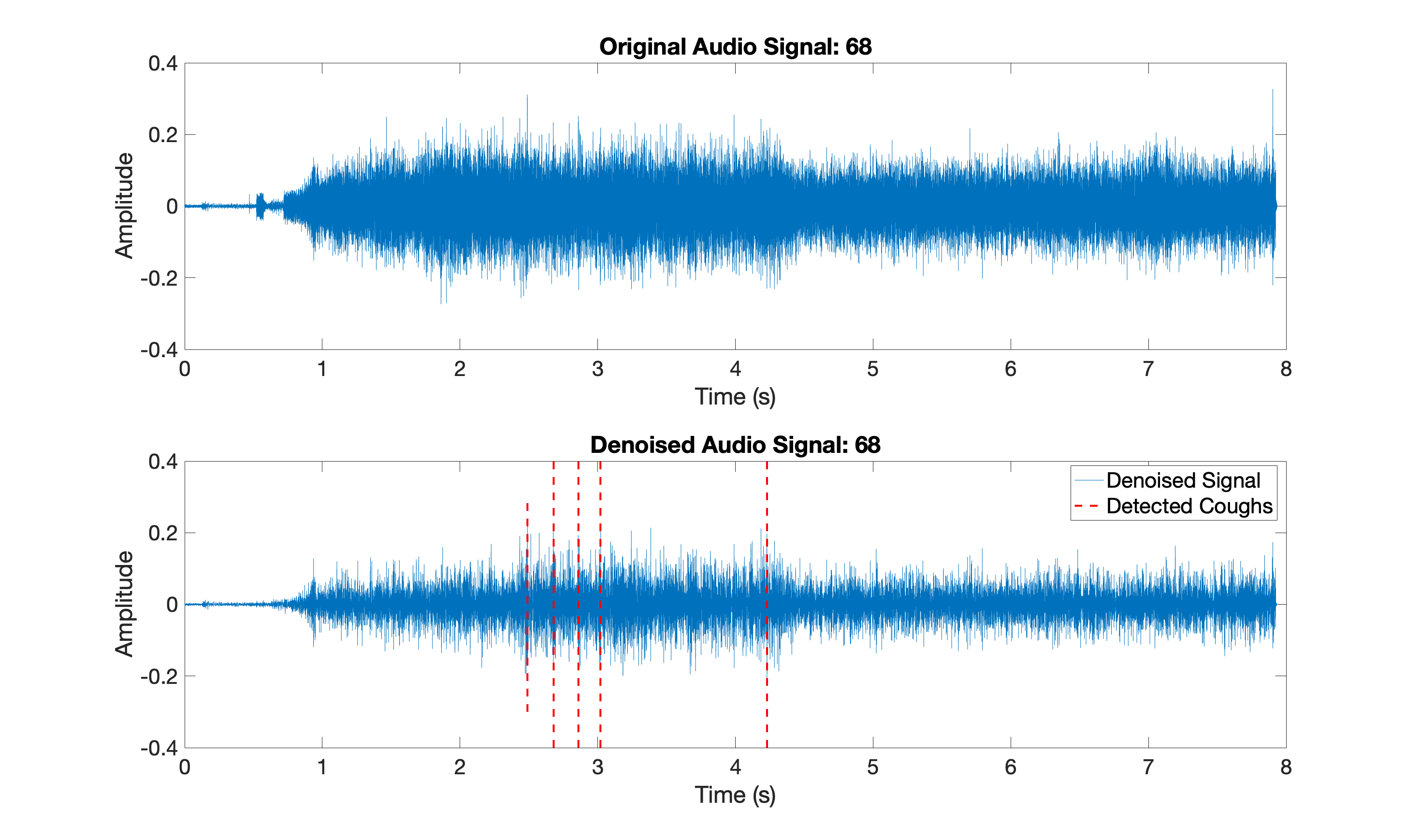}  
    \caption{Cough 0: False Positive}
    \label{No cough 68}
\end{subfigure}
\caption{Cough detection in CW2 devices.}
\label{CW2}
\end{figure}

Furthermore, as illustrated in \autoref{E70}, when using the E70 device, true positive cough detections (as shown in \autoref{Cough 10}) are characterised by high amplitude and closely spaced time frames, whereas falsely detected coughs (as depicted in \autoref{No cough 55}) exhibit lower amplitude. This suggests that while some low-energy events were detected, they are likely attributable to pressure signals from the machinery. \autoref{CW2} corroborates the findings presented in \autoref{Cough CaseStudy}, demonstrating higher amplitudes compared to \autoref{E70}. Similar patterns are observed for both true positives and false positives in \autoref{CW2}.

\begin{figure}[!htb]
\centering
    \makebox[\textwidth]{
        \includegraphics[width=\textwidth, height = 8cm]{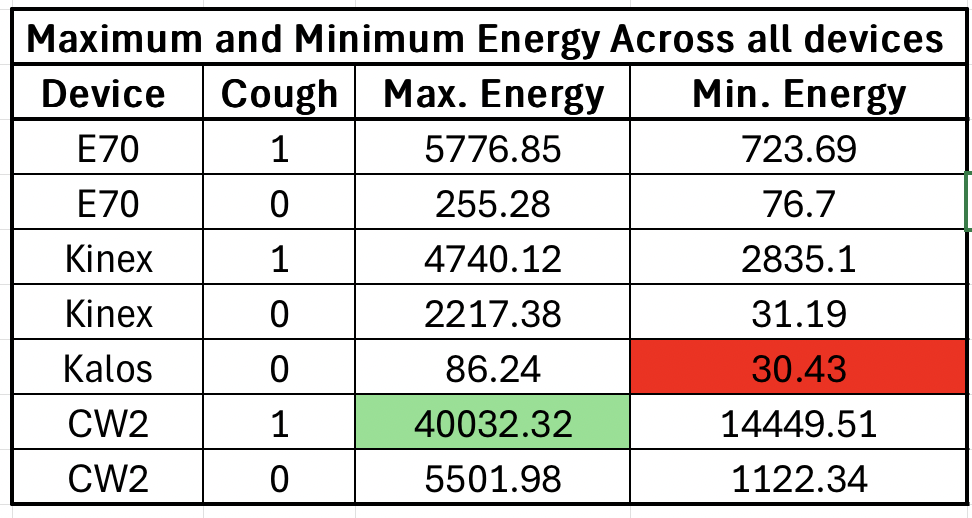}
    }
    \caption{Maximum and minimum energy levels across all 81 files for all devices, with maximum level highlighted in green and minimum level highlighted in red.}
    \label{MaxMin Energy}
\end{figure}

In a further assessment across all files and devices for both cough and non-cough events (see \autoref{MaxMin Energy}), CW2 consistently exhibited the highest energy levels during true positive cough detections, while the Kalos device consistently displayed the lowest energy levels during non-cough events. This finding highlights that, although some level of energy was detected during non-cough events (cough = 0), it was significantly lower than that of cough events (cough = 1) across all devices. However, the difference in energy levels between cough and non-cough events for the Kinex device was less pronounced compared to the E70 and CW2 devices.

\subsection{Results 3 - Machine Learning}
\label{sec:Results 3}
The results of the prediction are shown in the table \ref{tab: scores}. The accuracy and precision scores for cough predictions demonstrate that the machine learning model is able to predict with a good degree of accuracy and precision. 
\begin{table}[h!]
    \centering
    \begin{tabular}{|c|c|}
        \hline
        \textbf{Score} & \textbf{Value} \\ 
        \hline
        \textbf{Accuracy}&  \( 93.0 \pm 15\% \)\\
        \hline
        \textbf{Precision}&  \( 99.0 \pm 4.0\% \)\\
        \hline
    \end{tabular}
    \caption{Mean Accuracy and Precision Scores over 20 predictions using random training sets}
    \label{tab: scores}
\end{table}

\subsection{Results 4 - Wavelet Denoising}
\label{sec:results 4}
Table \ref{tab:cough_detect} shows the confusion matrix for the classification. Note that only 80 recordings were analysed: one recording (\texttt{52.wav}) was excluded, as it could not be imported using the \texttt{scipy.io.wavfile} library.

\begin{table}[ht]
    \centering
    \begin{tabular}{cc|cc|l}
    \cline{3-4}
                                                                            &                                                               & \multicolumn{2}{c|}{Predicted}                                                                                                  &  \\ \cline{2-4}
    \multicolumn{1}{c|}{}                                                   & \begin{tabular}[c]{@{}c@{}}Total \\ 39 + 41 = 80\end{tabular} & \multicolumn{1}{c|}{\begin{tabular}[c]{@{}c@{}}Cough\\ 24\end{tabular}} & \begin{tabular}[c]{@{}c@{}}No cough\\ 56\end{tabular} &  \\ \cline{1-4}
    \multicolumn{1}{|c|}{\multirow{2}{*}{\rotatebox[origin=c]{90}{Actual}}} & \begin{tabular}[c]{@{}c@{}}Cough\\ 39\end{tabular}            & \multicolumn{1}{c|}{15}                                                 & 24                                                    &  \\ \cline{2-4}
    \multicolumn{1}{|c|}{}                                                  & \begin{tabular}[c]{@{}c@{}}No cough\\ 41\end{tabular}         & \multicolumn{1}{c|}{9}                                                  & 32                                                    &  \\ \cline{1-4}
    \end{tabular}
    \caption{Confusion matrix of the spectral analysis-based cough classifier over the whole dataset.}
    \label{tab:cough_detect}
\end{table}

Of the recordings analysed, 47  (59\%) were correctly classified. The large error rate was predominantly caused by under-detecting coughs: though 41 recordings (51\%) contained coughs, the classifier only marked 24 recordings (30\%) as containing coughs, 9 of which were false positives. This under-detection could be caused by the classification applied: single cough sounds may be missed by requiring 2 transients to occur in the 2s following a pressure change. There may also be signal noise due to excessive spectral denoising. Some MI-E devices use brushed motors, and as such make a rhythmic sound rather than a static noise - in these cases, the assumptions made to apply spectral subtraction no longer apply. Further, full-spectrum transients may not be present in some of the recordings if lower frequencies are over-represented, which would occur if the phone or MI-E device was placed in a corner of the room \cite{oberhettinger1958diffraction}.

There was one subset of the recordings (\texttt{1.wav}-\texttt{16.wav}) for which this analysis performed well, correctly identifying 14/16 coughs (88\%). The confusion matrix for this subset is given in Table \ref{tab:cough_detect_2}. This group of recordings all used the same MI-E device (E-70), and had high cough peak flow (CPF) relative to the wider dataset. It is possible that the coughs were most prominent in this group of recordings, meaning there was more residual signal once the spectral denoising had been applied.

\begin{table}[ht]
    \centering
    \begin{tabular}{cc|cc|l}
    \cline{3-4}
                                                                            &                                                              & \multicolumn{2}{c|}{Predicted}                                                                                                 &  \\ \cline{2-4}
    \multicolumn{1}{c|}{}                                                   & \begin{tabular}[c]{@{}c@{}}Total \\ 16 + 0 = 16\end{tabular} & \multicolumn{1}{c|}{\begin{tabular}[c]{@{}c@{}}Cough\\ 14\end{tabular}} & \begin{tabular}[c]{@{}c@{}}No cough\\ 2\end{tabular} &  \\ \cline{1-4}
    \multicolumn{1}{|c|}{\multirow{2}{*}{\rotatebox[origin=c]{90}{Actual}}} & \begin{tabular}[c]{@{}c@{}}Cough\\ 16\end{tabular}           & \multicolumn{1}{c|}{14}                                                 & 2                                                    &  \\ \cline{2-4}
    \multicolumn{1}{|c|}{}                                                  & \begin{tabular}[c]{@{}c@{}}No cough\\ 0\end{tabular}         & \multicolumn{1}{c|}{0}                                                  & 0                                                    &  \\ \cline{1-4}
    \end{tabular}
    \caption{Confusion matrix of the spectral analysis-based cough classifier over the E70 subset.}
    \label{tab:cough_detect_2}
    \end{table}

\subsection{Results 5 - Predictive Modelling}

The predictive model (\ref{eq:gov}) and boundary conditions (\ref{eq:bcs}) are solved for parameter choices listed in Table \ref{tab:paras}. In particular, we consider the role of the parameter $k$ which is proportional to the viscosity of the air flow.

\begin{figure}[!htb] 
\centering
\includegraphics[width=0.88\textwidth]{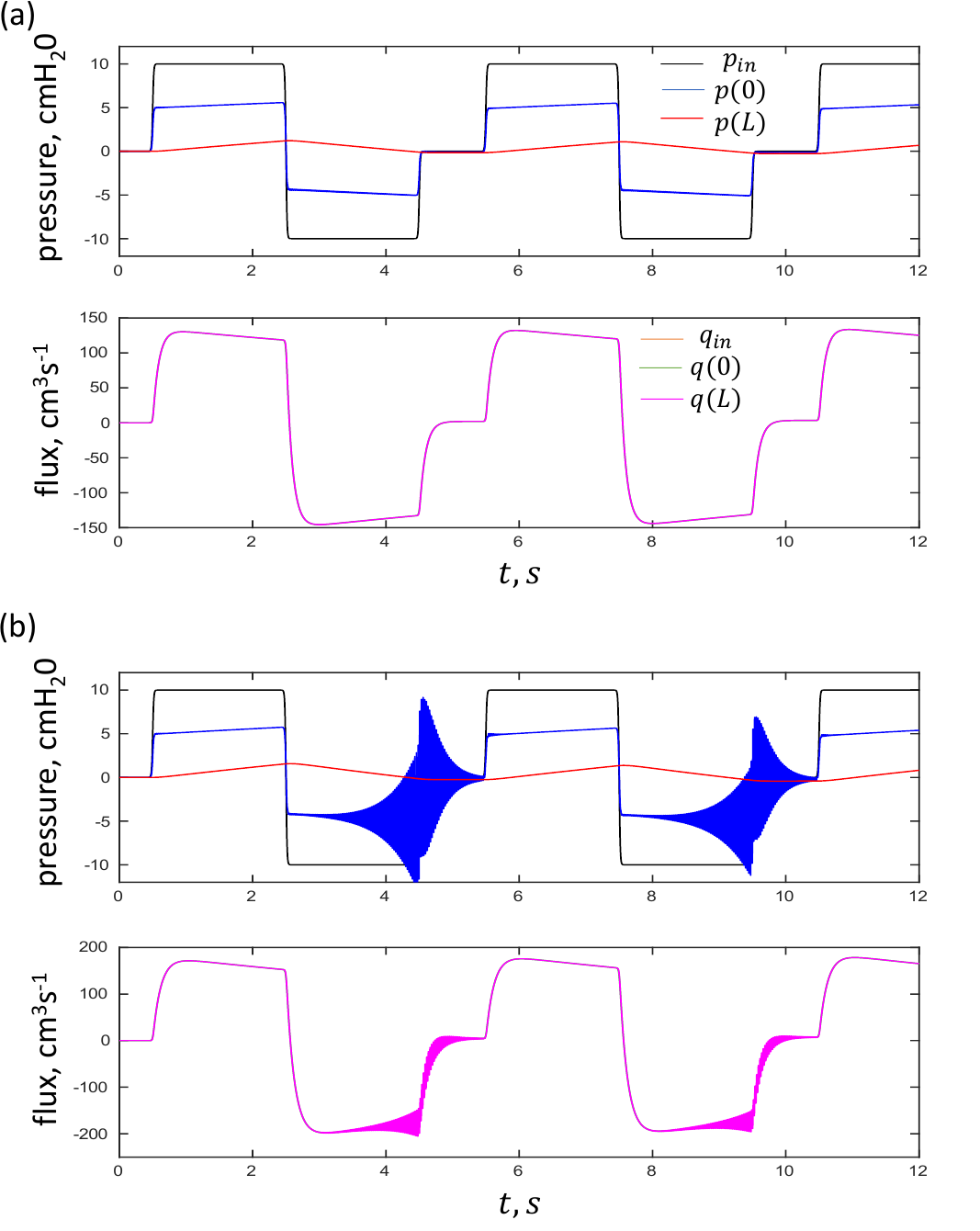} 
\caption{Time-traces of the pressure and flow rate at various locations along the lung system for: (a) $k=3000$; (b) $k=4000$. In each case the upper panel plots time-traces of the air pressure at the inlet to the mask (black line), inlet to the central airways (blue line) and the inlet to the peripheral airways (red line). The lower panel plots time-traces of the air flow rate at the inlet to the mask (orange line), the inlet to the central airways (green line) and the inlet to the peripheral airways (magenta line).}
\label{fig:pressure_time}
\end{figure}

For large viscosity ($k=4000$) the model predictions are summarised in Fig.~\ref{fig:pressure_time}(a), showing time-traces of the pressure in response to a prescribed (oscillating) input mimicking the mechanical insufflation machine (Fig.~\ref{fig:pressure_time}a, top) and a corresponding time-trace of the flow rate (Fig.~\ref{fig:pressure_time}a, lower). As the prescribed inlet pressure transitions to become positive, the corresponding air pressure along the compliant segment becomes approximately half the inlet value (Fig.~\ref{fig:pressure_time}a, top), while the flow rate increases more slowly than the pressure increase but eventually approaches a positive constant (air being drawn into the lung), indicating that flow is being drawn into the lung (Fig.~\ref{fig:pressure_time}a, lower). Over time the peripheral airways gradually expand and increase in pressure, while the central airway pressure also increases slowly and the flow rate gradually decreases. However, when the inlet pressure abruptly transitions to become negative the corresponding air pressure in the vessel decreases to a much lower value, which again decreases slowly as air leaves out of the peripheral airways again. As the pressure changes sign, the flow rate across the inlet abruptly reverses, and flow is now ejected from the lung. This process repeats, following the prescribed pressure profile of the device, where the central airway pressure is always approximately half of the prescribed inlet pressure.

However, the response is rather different when the friction factor is decreased ($k=3000$, Fig.~\ref{fig:pressure_time}b). While relatively unchanged during periods of positive input pressure, as the system abruptly transitions to negative inlet pressure and the central airway pressure rapidly decreases, this triggers a high-frequency fluttering instability in the central airways; this instability is evident on the time-traces of both the pressure (Fig.~\ref{fig:pressure_time}b, top) and the flow rate (Fig.~\ref{fig:pressure_time}b, lower), where the time-traces oscillate so quickly that the lines almost overlap on the scale plotted. This high-frequency oscillatory event is reminiscent of a dynamic cough, where the fluctuations lead to the generation of an audible signal. The violence of this coughing event becomes increasingly strong as the friction in the flow decreases (\emph{i.e.} the flow becomes increasingly inviscid), but it becomes increasingly difficult to resolve these events numerically.

\section{Discussion} 
\label{sec:disc}
\subsection{Results 1 - Adaptive Neuro-Fuzzy Inference System (ANFIS)}
Method focused on developing a robust method for detecting respiratory abnormalities, specifically cough events, by leveraging a combination of Mel Frequency Cepstral Coefficients (MFCCs) and Melspectrogram features. This combined feature set was used to train an Adaptive Neuro-Fuzzy Inference System (ANFIS) to classify respiratory sounds in denoised audio signals. The experimental results demonstrated that the integration of these two feature extraction techniques provides a more comprehensive representation of the audio signal, capturing both temporal and spectral characteristics.

The performance evaluation, conducted using 5-fold cross-validation, showed promising results with a mean validation accuracy of 90.16\% and a standard deviation of \(\pm 7.10\%\). This indicates that the model is capable of generalizing well across different data subsets. The ANFIS model’s ability to incorporate fuzzy logic, alongside its neural network architecture, enabled it to handle uncertainties and nonlinearities in the respiratory sound patterns, making it a suitable choice for this task.

However, some limitations were observed during the analysis. For instance, the model showed signs of overfitting in some folds, particularly in those with fewer examples of specific sound events. This suggests that the model may benefit from additional regularization techniques or the use of a larger and more diverse dataset. Additionally, the computational complexity of ANFIS increases significantly with the size of the dataset, which could become a bottleneck when scaling up to larger datasets.

\subsection{Results 2 - Wavelet Denoising}
Based on the analysis of denoised audio signals using energy-based spectral analysis for cough detection, several key observations have been identified. The CW2 device consistently showed the highest energy levels during true positive cough detections across various files and devices, highlighting its effectiveness in identifying cough events. In contrast, the Kalos device consistently exhibited the lowest energy levels during non-cough events, indicating its capability to distinguish background noise from cough sounds. These findings suggest that combining spectral analysis with denoising techniques can enhance cough detection by utilising energy thresholds in spectrograms. However, the study has limitations, including variations in ambient noise levels and the challenge of precisely defining cough events based solely on energy levels.
\subsection{Results 3 - Machine Learning}
Based on the results, it can be concluded that the ML approach is suitable for detecting coughs with the level of accuracy and precision demonstrated in the results. The accuracy shows a SD of \(\pm 15\%\), because three of the randomly selected training sets resulted in lower accuracy levels. One of the limitations of this model is that the data-set is small with 81 signals. With a larger data-set of 1000+ audio signals, the training would be more robust and one can create multiple validation and test data-sets to ensure better generalisability. However, this approach did not work for detecting leaks and more work is required ton Feature Engineering to generate s suitable data-set.

\section{Next steps}

Future work will focus on refining thresholding algorithms, exploring machine learning approaches to improve classification accuracy, and validating the method across diverse real-world environments to enhance the reliability of automated cough detection systems. Potential pathways for enhancement include:

\begin{itemize} \item \textbf{Energy and Spectral Features:} Features derived from spectrograms, such as energy distribution, spectral centroid, and bandwidth, can provide discriminative information for distinguishing between cough and non-cough events.

\item \textbf{Neural Networks:} Deep learning models, such as Convolutional Neural Networks (CNNs), can automatically learn relevant features from spectrograms or time-frequency representations of audio signals for more accurate cough detection.

\item \textbf{Support Vector Machines (SVMs):} SVMs can be trained to classify spectrogram features extracted from audio signals into cough and non-cough categories.
\end{itemize}

\begin{figure}[h]
    \centering
    \includegraphics[width=\textwidth]{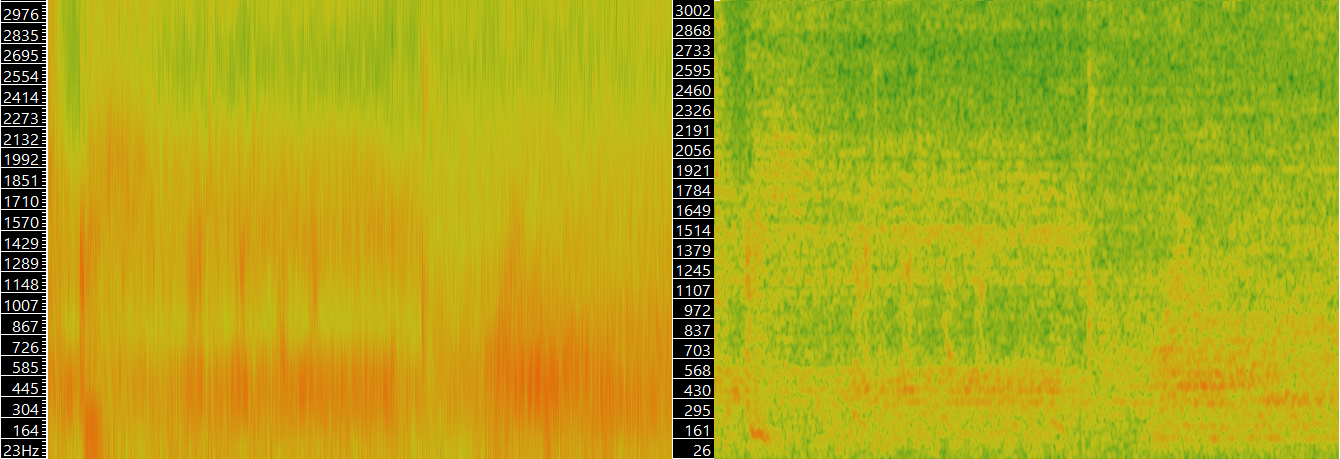}
    \caption{Two spectrograms of the same recording, with a window size of 256 samples (left) and 2048 samples (right). Smaller window sizes limit frequency resolution, resulting in vertical smearing (left). Larger window sizes limit time resolution, resulting in horizontal smearing.}
    \label{fig:spect_window}  
\end{figure}

There are many ways this approach could be improved, to perform better on the wider dataset or other recordings. 
Firstly, the spectral denoising approach could be improved by tuning the parameters for the \texttt{noisereduce} algorithm, and isolating noise snippets from each recording. If spectral gating can provide more noise reduction, then the spectral subtraction step can be skipped to avoid removing signal.
With more time, the locations of detected transients could be passed as inputs to the machine learning approach described in \ref{sec:method 3}. The transient portions of the recordings could be extracted, and these used as additional inputs.
Further, much could be done to better explore the full frequency range of the recordings in more detail. The default window size of the STFT transform used in this section is 256 samples. There is a tradeoff between time and frequency resolution determined by the window size, as shown in Fig \ref{fig:spect_window}.
Using a larger window size would permit better analysis of lower frequencies. Multiple STFTs could be computed with different window sizes, so that different aspects of coughs can be captured by analysing each.

The theoretical model constructed and solved above is sufficient to predict a high-frequency fluttering event reminiscent of a cough provided the friction factor is sufficiently low. Within this model it will be possible to:
\begin{itemize}
\item systematically explore the parameter space to predict when coughs are likely to occur;
\item predict the corresponding acoustics of the cough, to be directly compared to the acoustic measurements analysed in the rest of this report;
\item include the effect of leakiness in the mask region, to quantify its effect on the acoustic signal and the effectiveness of the pressure ramping protocol.
\end{itemize}

\section*{Author Contributions}
\addcontentsline{toc}{section}{Author Contributions}
Jonathan Brady, Vijay Chandiramani, Emmanuel Lwele, Aminat Yetunde Saula and Peter Stewart contributed equally to the preceding analysis and authorship of this report. The challenge holders, Michelle Chatwin and Toby Stokes, contributed numerous insights and data, and their contributions are greatly appreciated.

\label{EndOfText}
\newpage
\pagenumbering{Roman} 
\addcontentsline{toc}{section}{List of Figures}
\fancyfoot[C]{Page \thepage\ of \pageref{endOfDoc}}
\listoffigures
\thispagestyle{fancy}


\newpage
\addcontentsline{toc}{section}{References}
\bibliography{main.bib}
\bibliographystyle{ieeetr}

\label{endOfDoc}
\end{document}